\documentclass[11pt]{article}
\usepackage{amsmath,amssymb}
\usepackage[pdftex]{graphicx}
\pdfoutput=1
\usepackage{float}

\makeatletter
\renewcommand{\section}{%
 \@startsection{section}{1}{\z@}{-3.5ex \@plus -1ex \@minus -.2ex}{2.3ex \@plus.2ex}%
  {\normalfont\large\bfseries}}
\makeatother

\makeatletter
\renewcommand{\subsection}{%
 \@startsection{subsection}{1}{\z@}{-3.5ex \@plus -1ex \@minus -.2ex}{2.3ex \@plus.2ex}%
  {\normalfont\normalsize\bfseries}}
\makeatother

\makeatletter
\renewenvironment{thebibliography}[1]
{\section*{\refname\@mkboth{\refname}{\refname}}%
   \list{\@biblabel{\@arabic\c@enumiv}}%
        {\settowidth\labelwidth{\@biblabel{#1}}%
         \leftmargin\labelwidth
         \advance\leftmargin\labelsep
	 \setlength\itemsep{+2pt}%
	 \setlength\baselineskip{11pt}%
         \@openbib@code
         \usecounter{enumiv}%
         \let\p@enumiv\@empty
         \renewcommand\theenumiv{\@arabic\c@enumiv}}%
   \sloppy
   \clubpenalty4000
   \@clubpenalty\clubpenalty
   \widowpenalty4000%
   \sfcode`\.\@m}
  {\def\@noitemerr
    {\@latex@warning{Empty `thebibliography' environment}}%
   \endlist}
\makeatother

\def\slashchar#1{\setbox0=\hbox{$#1$}
\dimen0=\wd0 
\setbox1=\hbox{/} \dimen1=\wd1 
\ifdim\dimen0>\dimen1 
\rlap{\hbox to \dimen0{\hfil/\hfil}} 
#1 
\else 
\rlap{\hbox to \dimen1{\hfil$#1$\hfil}} 
/ 
\fi}

\allowdisplaybreaks[4] 

\renewcommand{\thefootnote}{\fnsymbol{footnote}}

\begin{document}

\title{{\Large Quark pair creation in color electric fields and effects of magnetic fields} } 
\author{{\normalsize Naoto Tanji}\footnote{\textit{E-mail address}: tanji@nt1.c.u-tokyo.ac.jp } }
\date{ \normalsize{\textit{Institute of physics, University of Tokyo, Komaba, Tokyo 153-8902, Japan} } } 

\maketitle

\renewcommand{\thefootnote}{$*$\arabic{footnote}}

\begin{abstract}
The time evolution of a system where a uniform and classical SU(3) color electric field 
and quantum fields of quarks are dynamically coupled with each other
is studied focusing on non-perturbative pair creation and its back reaction. 
We characterize the color direction of the electric field in a gauge invariant way, and investigate its dependence. 
Momentum distributions of created quarks show plasma oscillation as well as quantum effects 
such as the Pauli blocking and interference.
Pressure of the system is also calculated, and we show that pair creation moderates degree of anisotropy of pressure.  
Furthermore, enhancement of pair creation and induction of chiral charge 
under a color magnetic field which is parallel to the electric field are discussed. 
\end{abstract}

\vspace{-290pt}
\begin{flushright}
UT-Komaba/10-1
\end{flushright}
\vspace{250pt}

\section{Introduction} \label{sec:intro}
Study of non-perturbative pair creation from a classical electric field, 
which is known as the Schwinger mechanism \cite{Schwinger}, has a long history and wide range of applications 
(see Ref. \cite{Ruffini} for a recent review). 
One of those applications can be found in studies of relativistic heavy-ion collisions, where
the Schwinger mechanism has been used as a mechanism of matter formation from a color flux tube \cite{Gatoff}.
The color flux-tube model assumes that a strong color electric field is formed in a beam direction 
just after two nuclei collide and pass through each other \cite{Low,Nussinov}. 
Formation of longitudinal color electric fields is also predicted in the framework of color glass condensate 
\cite{Kovner,Lappi}. 
Therefore, particle production due to the Schwinger mechanism attracts renewed interest 
\cite{Kharzeev2005,Kharzeev2007a,Castorina,Fukushima2009,Levai-Skokov}. 

Under these circumstances, getting an understanding of how an initial electric field and created particles evolve in time 
is of prime importance.
To properly describe the time evolution, calculating vacuum persistence probability or pair creation probability, 
which were first derived by Schwinger, is not sufficient \cite{Tanji}, and
an electric field should be treated as a dynamical variable rather than
a background field controlled by hand, i.e. back reaction should be taken into account. 
There have been considerable numbers of studies treating back reaction; the ones based on a kinetic theory 
\cite{Gatoff,Asakawa,Dawson}
and the others on quantum field theory \cite{Tanji,Kluger1991-1993,Kluger1998,Bloch,Vinnik}. 
To our knowledge, however, field theoretical treatment of the back reaction problem
under a \textit{color} electric field has been lacking. 
Therefore, in this paper we investigate the pair creation of quarks under a color electric field incorporating back reaction. 

In studies of physics under non-Abelian electromagnetic fields, SU(2) theory has been often used for simplicity. 
In the case of SU(3), however, a new feature arises: anisotropy in color space. 
It has been shown that an SU(3) color electric field has two independent directions 
and it is characterized by two gauge invariant parameters:
one of them is determined by its field strength and the other is related with the color direction of the field
\cite{Nayak2005a,Nayak2005b}. 
More generally, an SU($N_c$) color vector has $(N_c -1)$-independent directions in color space, 
and physical contents can generally depend on a color direction of an electric field \cite{Ambjorn1979}. 
In this paper, we deal with SU(3) color electric fields and examine the color direction dependence. 

Not only new features which arise in non-Abelian fields, we also analyze phenomena 
whose essence is common to the Abelian case. 
Collective motion of created particles which couples to an electric field shows plasma oscillation. 
During this evolution, several phenomena are observed: 
suppression of pair creation or annihilation of the particles due to the Pauli blocking, 
damping of the electric field, 
and rapid oscillations in the momentum distribution of the created particles due to interference. 
We shall give an analysis of these phenomena to advance an understanding of physics in pair creation. 

We take a uniform color electric field as an initial state. 
Pressure of this initial state is quite anisotropic: 
the longitudinal pressure is negative and the transverse pressure is positive. 
Therefore, if local thermalization is achieved starting from the flux-tube initial condition,
isotropization of pressure should be needed during the time evolution. 
However, the full understanding of a thermalization process in heavy-ion collisions has not been obtained. 
In this paper, we examine the role of pair creation for the isotropization of pressure as a first step to 
understand a mechanism of thermalization in heavy-ion collisions. 

One of remarkable differences of the color flux tube given by the color glass condensate from that in the original
flux-tube model is
the existence of a longitudinal color magnetic field in addition to an electric field \cite{Lappi}. 
It has been shown that a longitudinal magnetic field enhances pair creation of fermions and
speeds up the decay of an electric field in the previous paper \cite{Tanji}. 
We extend it to the quark pair creation under a longitudinal color electric and magnetic field. 

Furthermore, we study induction of chiral charge due to pair creation under a magnetic field. 
Since the chiral anomaly is a semi-classical effect where the quantum aspect of a gauge field is unnecessary,
we can also apply our framework to study the chiral anomaly due to pair creation. 
The relation between pair creation and the chiral anomaly has been also studied 
in Refs. \cite{Ambjorn1983,Iwazaki2009}. 
Emergence of a nonzero chirality in heavy-ion collisions attracts interest in the context of 
the chiral magnetic effect \cite{Kharzeev2007b}. 

The remainder of this paper is organized as follows.
In the next section, we shall explain the Abelianization of a color electromagnetic field, 
and introduce the parameter characterizing the color direction of the field. 
Although this formalism is essentially the same as that given in Ref. \cite{Gyulassy-Iwazaki}, 
we make the existence of color direction dependence clearer with the help of the method in 
Refs. \cite{Nayak2005a,Nayak2005b}.  
In Section \ref{sec:canonical}, we introduce time-dependent particle picture 
to describe the time evolution of the system. 
Then, we shall show our numerical results in Section \ref{sec:ele}. 
Time evolution of momentum distribution functions of created quarks, color current density, electric field strength 
and pressure of the system are displayed and discussed. 
Color direction dependence of the results is also examined there. 
In Section \ref{sec:mag}, effects of a longitudinal magnetic field, i.e. enhancement of pair creation
and induction of chiral charge, are discussed. 

\section{General framework} \label{sec:frame}
Quark pair creation incorporated with back reaction is described by the following Lagrangian density
\begin{equation}
\mathcal{L} = \bar{\psi } \left( i\gamma ^\mu D_\mu -m\right) \psi 
              -\frac{1}{4} F_{\mu \nu } ^a F^{a \mu \nu} , \label{lag}
\end{equation}
where $\psi $ is a quark field
and color indices $i \ (i=1,2,\cdots ,N_c )$ are omitted.  
We assume for simplicity that each flavor has the same mass $m$, and flavor indices are also omitted. 
The number of flavor is set to be $N_f =3$ throughout this paper.

The covariant derivative and the field strengths are defined in terms of a background gauge field $A_\mu ^a$ as
\begin{gather}
D_\mu = \partial _\mu +igT^a A_\mu ^a, \label{coderi1} \\
F_{\mu \nu } ^a = \partial _\mu A_\nu ^a -\partial _\nu A_\mu ^a -gf^{abc} A_\mu ^b A_\nu ^c , \label{strength}
\end{gather}
where $T^a$ is the generator in the fundamental representation of gauge group SU($N_c $),
and $f^{abc} $ is the anti-symmetric structure constant ($a,b,c=1,2,\cdots ,N_c ^2 -1$).

The equations of motion for $A_\mu ^a $ and $\psi $ now read
\begin{gather}
D_\mu F^{a \mu \nu} = g \bar{\psi } \gamma ^\nu T^a \psi, \label{Maxwell1} \\
\left( i\gamma ^\mu D_\mu -m\right) \psi = 0 . \label{Dirac1}
\end{gather}
Because we treat the gauge field $A_\mu ^a$ as a classical background field, the charge current operator
$g \bar{\psi } \gamma ^\nu T^a \psi $ in Eq.\eqref{Maxwell1} is replaced by its expectation value
$g \langle \bar{\psi } \gamma ^\nu T^a \psi \rangle $ in the following. 
These coupled equations govern the pair creation and its back reaction.

We restrict the background field to spatially homogeneous and Abelian-like one, which is expressed as
\begin{equation}
F_{\mu \nu } ^a = \bar{F} _{\mu \nu } n^a , \label{Fna}
\end{equation}
where $\bar{F}_{\mu \nu } $ is an Abelianized field strength and is independent of the space coordinates.
$n^a $ is a constant vector indicating a color orientation of the electromagnetic field,
and is normalized so that $n^a n^a =1$. 
This field strength is given by the gauge field 
\begin{equation}
A_\mu ^a = \bar{A} _\mu n^a . \label{Ana}
\end{equation}
The relation between $\bar{F}_{\mu \nu } $ and $\bar{A} _\mu $ is just the same as that in the Abelian case:
\begin{equation}
\bar{F}_{\mu \nu } = \partial _\mu \bar{A} _\nu -\partial _\nu \bar{A} _\mu .
\end{equation}

Under the Abelianized gauge field \eqref{Ana}, the covariant derivative \eqref{coderi1} reads 
$D_\mu = \partial _\mu +ign^a T^a \bar{A}_\mu $. 
Because $\left( n^a T^a \right) _{ij} $ is an $N_c \times N_c $ hermitian matrix, 
we can diagonalize it by a unitary transformation\footnote{Diagonalization itself is always possible 
by local gauge transformation even if a color electric field is not spatially uniform.
In that case, however, $w_i$ acquire space dependence, and the method used in this paper 
is no longer applicable.}(global gauge transformation): 
$U n^a T^a U^\dagger = \mathrm{diag} ( w_1 ,\cdots ,w_{N_c } )$. 
Then, the equations of motion \eqref{Maxwell1} and \eqref{Dirac1} can be rewritten as
\begin{gather}
\partial _\mu \bar{F} ^{\mu \nu } = \sum _{i=1} ^{N_c } w_i g \langle \bar{\psi } ^i \gamma ^\nu \psi ^i \rangle ,
\label{Maxwell2} \\
\left[ i\gamma ^\mu (\partial _\mu +iw_i g \bar{A} _\mu ) -m\right] \psi ^i = 0 
 \hspace{20pt} (i=1,2,\cdots ,N_c ). \label{Dirac2}
\end{gather}
These equations have the same form with those in the Abelian theory, except the presence of 
the weight vector  $w_i $. 
Hence, we can solve the coupled equations \eqref{Maxwell2} and \eqref{Dirac2} as in the Abelian case \cite{Tanji}.
Each quark field $\psi ^i $ couples with the background field $\bar{A} _\mu $ via a coupling constant $w_i g$. 

\begin{figure}
\begin{center}
\includegraphics[scale=0.8]{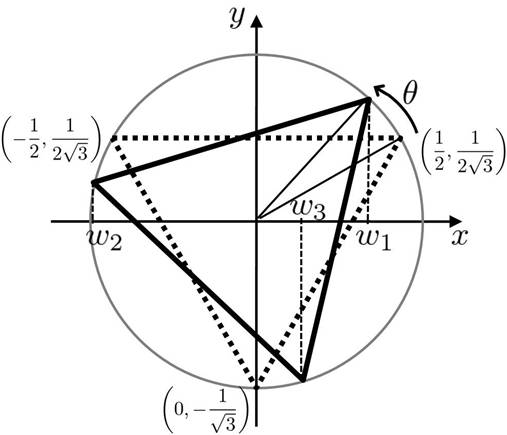} \\ 
\end{center}
 \vskip -\lastskip \vskip -3pt
\caption{A graphical representation of Eqs.\eqref{weight}.
The triangle of dotted lines corresponds to $\theta =0$ and is an usual weight diagram for a fundamental 
representation of SU(3).
Each $w_i $ is given by the $x$ coordinate of each vertex of the $\theta $-rotated triangle. }
\label{fig:triangle}
\end{figure}

In the case of SU(3), a diagonalized $n^a T^a $ may be expressed as
\begin{equation}
U n^a T^a U^\dagger = T^3 \cos \theta -T^8 \sin \theta , \label{UnTU}
\end{equation} 
because $T^3 =\mathrm{diag} (1/2,-1/2,0)$ and $T^8 =\mathrm{diag} (1/2\sqrt{3} ,1/2\sqrt{3} ,-1/\sqrt{3} )$ 
are diagonal.
(We represent $T^a $ as $T^a =\frac{1}{2} \lambda _a $ where $\lambda _a $ is the Gell-Mann matrix.) 
Then, $w_i $ are expressed in terms of $\theta $ as follows
\begin{equation}
w_1 = \frac{1}{\sqrt{3} } \cos (\theta +\frac{\pi }{6} ) , \
w_2 = \frac{1}{\sqrt{3} } \cos (\theta +\frac{5\pi }{6} ) , \
w_3 = \frac{1}{\sqrt{3} } \cos (\theta +\frac{3\pi }{2} ) . \label{weight}
\end{equation} 

The angle $\theta $ is related with the Casimir (gauge) invariant \cite{Nayak2005a,Nayak2005b}. 
In our parametrization\footnote{The relation between $\theta$ in this paper and those in Refs.\cite{Nayak2005a,Nayak2005b} is 
$\theta = \theta _\text{Ref.\cite{Nayak2005b}} +\frac{\pi }{6} 
= \frac{1}{2} \theta _\text{Ref.\cite{Nayak2005a}} +\frac{\pi }{6}$. }, 
the relation is 
\begin{equation}
\sin ^2 3\theta = 3C_2 ,
\end{equation}
where $C_2 = [d^{abc} n^a n^b n^c ]^2 $ is the second Casimir invariant for SU(3) and characterizes a direction
of the color electromagnetic field in a gauge invariant way. 

Fig.\ref{fig:triangle} is a graphical representation of Eqs.\eqref{weight}. 
Each $w_i $ is given by the $x$ coordinate of each vertex of the rotated triangle (weight diagram).  
Owing to the symmetry of the diagram, it is sufficient to take the angle $\theta $ restricted to 
$0\leq \theta \leq \frac{\pi }{6} $. 

\section{Canonical quantization in background fields} \label{sec:canonical}
To describe the time evolution of the system, 
we introduce an instantaneous particle picture and quantize the field using that particle picture. 
Of course, a definition of particle in the presence of a pair-creating background is rather ambiguous. 
Thus, the instantaneous particle picture should be regarded as our working definition to describe the system evolution.
To calculate field quantities such as current and energy density, we need to identify and subtract the contribution 
from the Dirac sea at each time. 
We can do this just by the normal ordering in terms of the instantaneous particle basis,
and then energy conservation is automatically guaranteed \cite{Tanji}. 
Therefore, one can interpret defining a particle picture at each time as a means to properly regularize field quantities,
and can consider particle number at intermediate time as a byproduct. 
Nevertheless, we will show time evolution of a particle number or momentum distribution function 
defined through the instantaneous particle picture in the following, 
because it behaves in a physically reasonable manner, and helps us understand the dynamics of this system. 

First, we consider only the electric fields. 
We will introduce the magnetic fields in Section \ref{sec:mag}. 
We set an initial state with no electric field and switch it on at time $t=0$. 
An initial field strength is $E_0$ and its direction is along the $z$-axis. 
The gauge $\bar{A} _0 = 0$ is chosen so that 
\begin{gather}
\bar{A} _3 (t) = 0 \ \ ( t\leq 0 ), \notag \\
\frac{d\bar{A} _3 }{dt} \bigg| _{t=0 } = -E_0 ,
\end{gather}
and $\bar{A} _1 = \bar{A} _2 =0$. 
After switching on, $\bar{A} _3 (t) $ evolves according to Eqs.\eqref{Maxwell2} and \eqref{Dirac2}.

A quantized quark field may be expanded as
\begin{equation}
\psi ^i (x) = \sum _{s=\uparrow ,\downarrow } \int \! d^3 p \left[
 {}_+ \! \psi ^{i\, \mathrm{in} } _{\mathbf{p} s} (x) a ^{i\, \mathrm{in} } _{\mathbf{p} ,s} + 
 {}_- \! \psi ^{i\, \mathrm{in} } _{\mathbf{p} s} (x) b^{i\, \mathrm{in} \dagger } _{-\mathbf{p} ,s} \right] ,
\end{equation}
where $a^{i\, \mathrm{in} } _{\mathbf{p} ,s} $ [$b^{i\, \mathrm{in} } _{\mathbf{p} ,s} $]
is the annihilation operator of a particle [antiparticle] with momentum $\mathbf{p} $ and spin-$s$ satisfying 
the anti-commutation relation 
$\{ a^{i\, \mathrm{in} } _{\mathbf{p} ,s} ,a^{j\, \mathrm{in} \dagger } _{\mathbf{q} ,s^\prime } \} 
 = \{ b^{i\, \mathrm{in} } _{\mathbf{p} ,s} ,b^{j\, \mathrm{in} \dagger } _{\mathbf{q} ,s^\prime } \} 
 = \delta _{ij} \delta _{ss^\prime } \delta ^3 (\mathbf{p} -\mathbf{q} ) $, and
${}_\pm \! \psi ^{i\, \mathrm{in} } _{\mathbf{p} s} (x) $ are classical solutions of the Dirac equation \eqref{Dirac2}. 
The superscript {\lq}in' distinguishes the initial condition for 
${}_\pm \! \psi ^{i\, \mathrm{in} } _{\mathbf{p} s} (x) $:
at $t<0$, ${}_+ \! \psi ^{i\, \mathrm{in} } _{\mathbf{p} s} (x) $ 
[${}_- \! \psi ^{i\, \mathrm{in} } _{\mathbf{p} s} (x) $] is identical to 
the positive [negative] energy solution of the free Dirac equation. 
We set the state to be in-vacuum $|0,\mathrm{in} \rangle $, where no particle exists initially and which is defined by 
$a^{i\, \mathrm{in} } _{\mathbf{p} ,s} |0,\mathrm{in} \rangle 
 = b^{i\, \mathrm{in} } _{\mathbf{p} ,s} |0,\mathrm{in} \rangle = 0 $. 

At $t>0$, ${}_\pm \! \psi ^{i\, \mathrm{in} } _{\mathbf{p} s} (x) $ evolve under influence 
of the electric field and become superposition of a positive and negative energy (frequency) state.
To describe the pair creation dynamically, we introduce a time-dependent particle picture by decomposing the field operator
$\psi ^i (x) $ into positive and negative frequency instantaneously:
\begin{equation}
\psi ^i (t_0 ,\mathbf{x} ) = \sum _{s=\uparrow ,\downarrow } \int \! d^3 p \left[
 {}_+ \! \psi ^{i\, (t_0 ) } _{\mathbf{p} s} (x) a ^i _{\mathbf{p} ,s} (t_0 ) + 
 {}_- \! \psi ^{i\, (t_0 ) } _{\mathbf{p} s} (x) b^{i\, \dagger } _{-\mathbf{p} ,s} (t_0 ) \right] , \label{psi}
\end{equation}
where ${}_+ \! \psi ^{i\, (t_0 ) } _{\mathbf{p} s} (x) $ [${}_- \! \psi ^{i\, (t_0 ) } _{\mathbf{p} s} (x) $]
is a positive [negative] frequency solution of the Dirac equation
under the pure gauge $\bar{A} _3 = \bar{A} _3 (t=t_0 ) $. 
Instantaneous particle picture is defined by $a ^i _{\mathbf{p} ,s} (t ) $ and $b ^i _{\mathbf{p} ,s} (t ) $. 
Of course, $a ^i _{\mathbf{p} ,s} (t ) $ and $b ^i _{\mathbf{p} ,s} (t ) $ agree with
$a ^{i\, \mathrm{in} } _{\mathbf{p} ,s} $ and $b ^{i\, \mathrm{in} } _{\mathbf{p} ,s} $ at $t< 0 $, respectively. 
The particle picture at time $t$ and that of the in-state are related by the time-dependent Bogoliubov transformation: 
\begin{equation}
\begin{matrix}
a^i _{\mathbf{p} s} (t) = \alpha ^i _{\mathbf{p} s} (t) a^{i\, \mathrm{in} } _{\mathbf{p} s } 
   +\beta ^i _{\mathbf{p} s} (t) b^{i\, \mathrm{in} \dagger } _{-\mathbf{p} s} , \\[+3pt]
b^{i\, \dagger } _{-\mathbf{p} s} (t) 
  = \alpha ^{i\, *} _{\mathbf{p} s} (t) b^{i\, \mathrm{in} \dagger } _{-\mathbf{p} s} 
   -\beta ^{i\, *} _{\mathbf{p} s} (t) a^{i\, \mathrm{in} } _{\mathbf{p} s } ,
\end{matrix} \label{Bogo}
\end{equation}
of which coefficients satisfy $| \alpha ^i _{\mathbf{p} s} (t) | ^2 +| \beta ^i _{\mathbf{p} s} (t) | ^2 = 1$ 
and are given by
\begin{equation}
\begin{matrix}
\alpha ^i _{\mathbf{p} s} (t) \delta ^3 (\mathbf{p} -\mathbf{q} +w_i g\bar{A} _3 (t) )
 = \displaystyle{\int } \! d^3 x {}_+ \! \psi ^{i\, (t) \dagger } _{\mathbf{p} s} (x) 
   {}_+ \! \psi ^{i\, \mathrm{in} } _{\mathbf{q} s} (x) , \\[+3pt]
\beta ^i _{\mathbf{p} s} (t) \delta ^3 (\mathbf{p} -\mathbf{q} +w_i g\bar{A} _3 (t) )
 = \displaystyle{\int } \! d^3 x {}_+ \! \psi ^{i\, (t) \dagger } _{\mathbf{p} s} (x) 
   {}_- \! \psi ^{i\, \mathrm{in} } _{\mathbf{q} s} (x) .
\end{matrix}
\end{equation}

A quark pair distribution function is defined by
\begin{equation}
f^i _{\mathbf{p} s} (t) = \langle 0,\mathrm{in} |a^{i\, \dagger } _{\mathbf{p} s} (t)a^i _{\mathbf{p} s} (t)
                           |0,\mathrm{in} \rangle \frac{(2\pi )^3 }{V} 
 = \langle 0,\mathrm{in} |b^{i\, \dagger } _{-\mathbf{p} s} (t)b^i _{-\mathbf{p} s} (t)
                           |0,\mathrm{in} \rangle \frac{(2\pi )^3 }{V} , \label{distri}
\end{equation}
where $V$ is the volume of the system. 
With the help of Eqs.\eqref{Bogo}, we can rewrite $f^i _{\mathbf{p} s} (t) $ 
in terms of the Bogoliubov coefficients:
\begin{equation}
f^i _{\mathbf{p} s} (t) = | \beta ^i _{\mathbf{p} s} (t) | ^2 .  \label{distri2}
\end{equation}
The expectation of the charge current operator, which is regularized by the normal ordering is 
\begin{equation}
\begin{split}
j_z (t) 
 &= \sum _{i=1,2,3} \langle 0,\mathrm{in} |:w_i g \bar{\psi } ^i \gamma _3 \psi ^i : |0,\mathrm{in} \rangle \\
 &= 2N_f \sum _{i=1,2,3} \sum _{s=\uparrow ,\downarrow } w_i g\int \! \frac{d^3 p }{(2\pi )^3 } 
    \frac{p_z }{\omega _p } f^i _{\mathbf{p} s} (t)
    +2N_f \sum _{i=1,2,3} \sum _{s=\uparrow ,\downarrow } w_i g\int \! \frac{d^3 p }{(2\pi )^3 } 
    \frac{m_\mathrm{T} }{\omega _p } g^i _{\mathbf{p} s} (t), \label{current}
\end{split} 
\end{equation} 
where the transverse mass $m_\mathrm{T} =\sqrt{m^2 +p_x ^2 +p_y ^2 } $ and the anomalous distribution 
\begin{equation}
g^i _{\mathbf{p} s} (t) = \frac{1}{2} \langle 0,\mathrm{in} |\left[e^{-2i\omega _p t} b^i _{-\mathbf{p} s} a^i _{\mathbf{p} s} 
   +e^{2i\omega _p t} a^{i\, \dagger } _{\mathbf{p} s} b^{i\, \dagger } _{-\mathbf{p} s} \right] |0,\mathrm{in} \rangle 
   \frac{(2\pi )^3 }{V}
 = \mathrm{Re} \left[ e^{-2i\omega _p t} \alpha ^i _{\mathbf{p} s} (t)\beta ^i _{\mathbf{p} s} (t) \right] \label{anom}
\end{equation}
are introduced.\footnote{Oscillating factor $e^{-2i\omega _p t} $ in the anomalous distribution is canceled out in total
because both $\alpha ^i _{\mathbf{p} s} (t)$ and $\beta ^i _{\mathbf{p} s} (t)$ contain the factor $e^{i\omega _p t} $.} 
The first term of the last expression of Eq.\eqref{current} is a conduction current, 
which is caused by movement of real particles and
the second is a polarization current, which is generated by variation of electric dipole density. 

\section{Pair creation under electric fields} \label{sec:ele} 

\subsection{Plasma oscillation and Pauli blocking} \label{subsec:pp}
\begin{figure}[t]
 \begin{tabular}{cc}
  \begin{minipage}{0.5\textwidth}
   \begin{center}
    \includegraphics[scale=0.45]{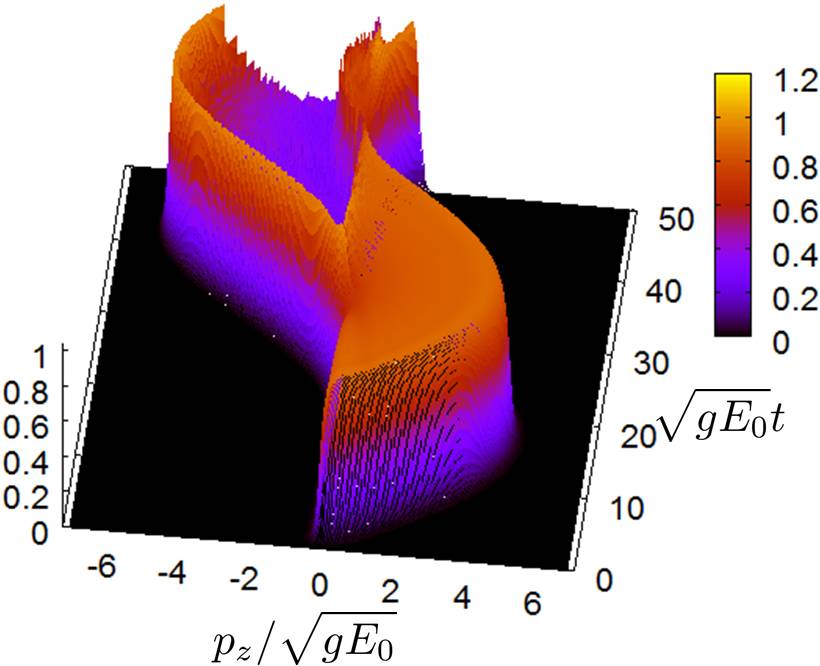} 
   \end{center}
  \end{minipage} & 
  \begin{minipage}{0.5\textwidth}
   \begin{center}
    \includegraphics[scale=0.45]{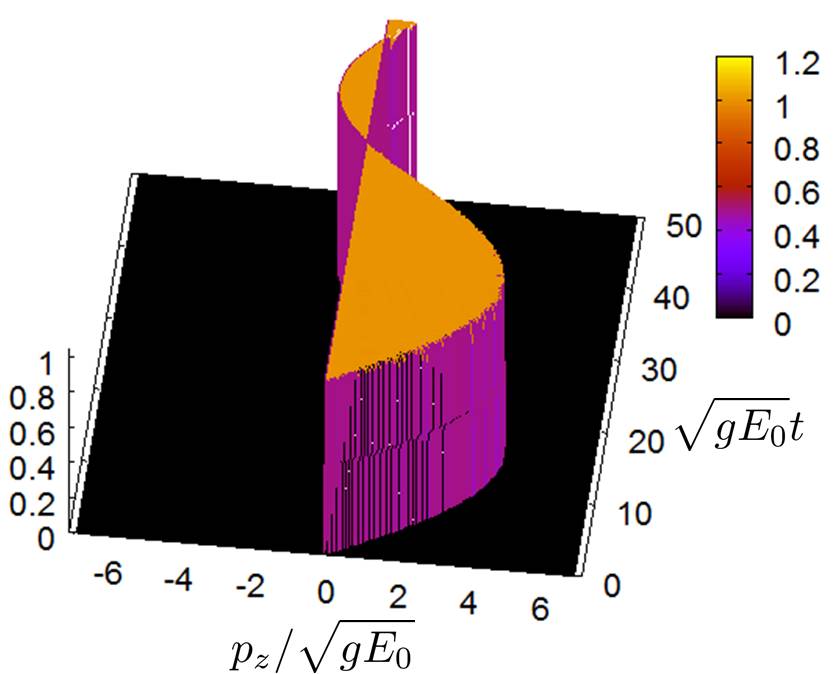} 
   \end{center}
  \end{minipage} \\
  (a) $a=\frac{m^2}{2gE_0 } =0.01 $ &
  (b) $a=0 $
 \end{tabular}
 \caption{Longitudinal momentum distributions of {\lq\lq}blue" quarks. $g=1,p_\text{T} =0,\theta =0 $.}
 \label{fig:Ldis1}
\end{figure}

\begin{figure}
 \begin{tabular}{cc}
  \begin{minipage}{0.5\textwidth}
   \begin{center}
    \includegraphics[scale=0.37]{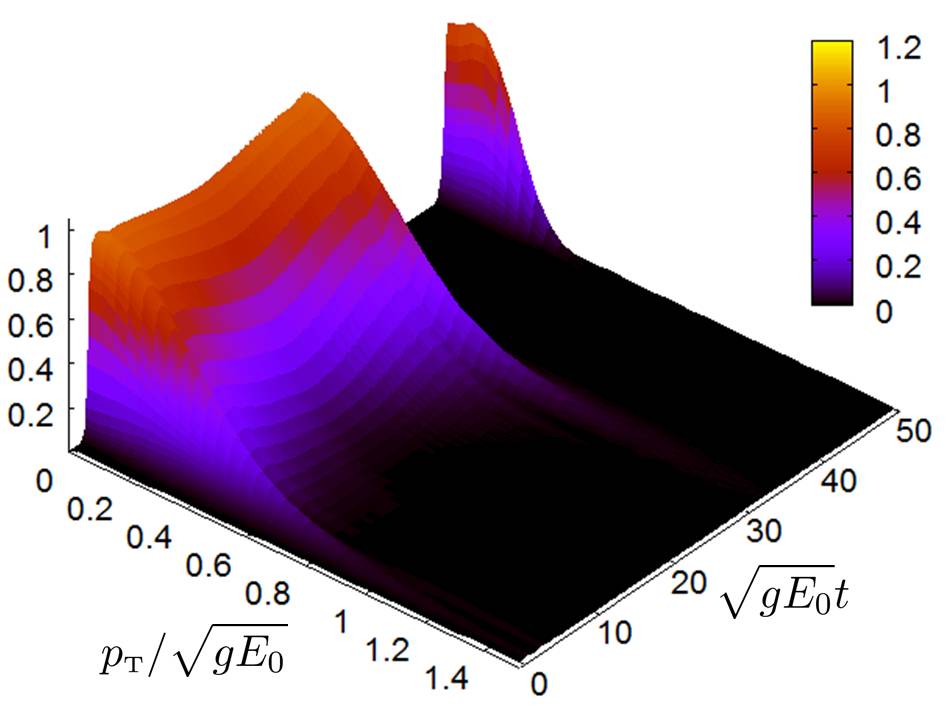} 
   \end{center}
  \end{minipage} & 
  \begin{minipage}{0.5\textwidth}
   \begin{center}
    \includegraphics[scale=0.37]{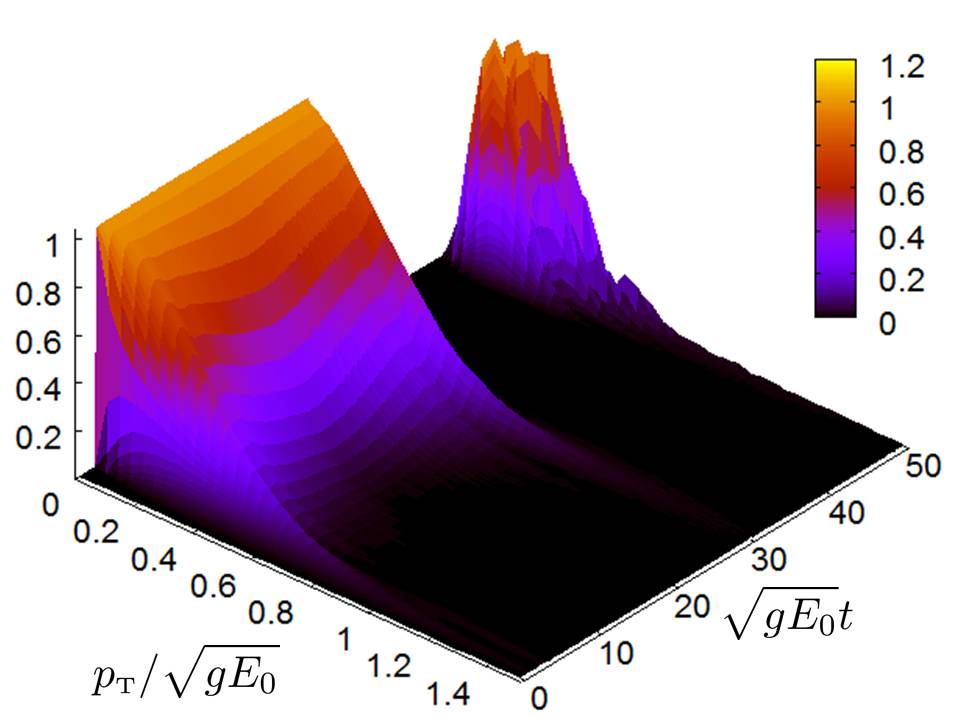} 
   \end{center}
  \end{minipage} \\
  (a) $a=\frac{m^2}{2gE_0 } =0.01 $ &
  (b) $a=0 $
 \end{tabular}
 \caption{Transverse momentum distribution of {\lq\lq}blue" quarks. $g=1, p_z /\sqrt{gE_0 } =1.0,\theta =0 $}
 \label{fig:Tdis}
\end{figure}

\begin{figure}
 \begin{tabular}{ccc}
  \begin{minipage}{0.33\textwidth}
   \begin{center}
    \includegraphics[scale=0.42]{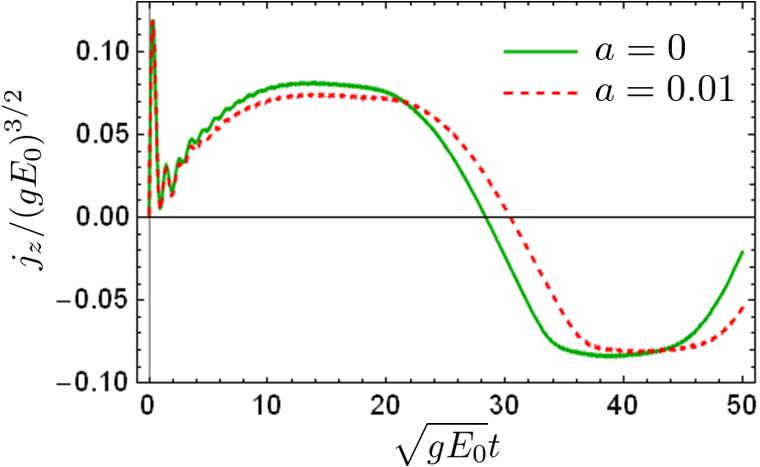} 
   \end{center}
  \end{minipage} & \hspace{-0.5cm}
  \begin{minipage}{0.33\textwidth}
   \begin{center}
    \includegraphics[scale=0.4]{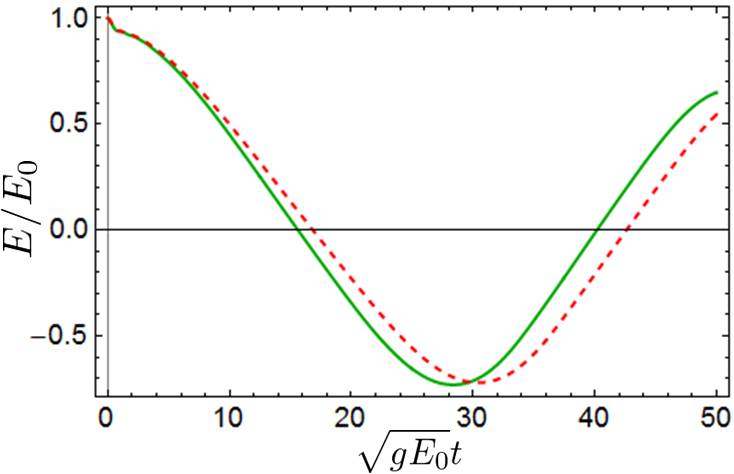} 
   \end{center}
  \end{minipage} & \hspace{-0.5cm}
  \begin{minipage}{0.33\textwidth}
   \begin{center}
    \includegraphics[scale=0.4]{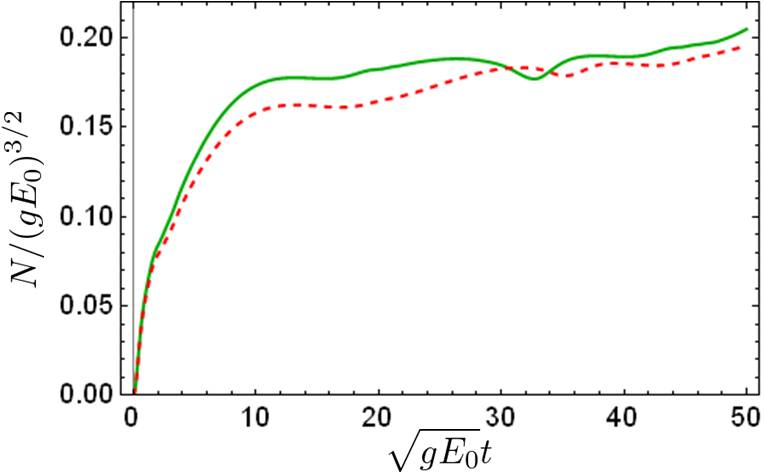} 
   \end{center}
  \end{minipage} \\
  (a) color current density &
  (b) electric field &
  (c) quark number density
 \end{tabular}
 \caption{Time evolution of the color current density, the electric field and quark number density.
          $g=1,\theta =0$.}
 \label{fig:current1}
\end{figure}

We have solved the coupled equation \eqref{Maxwell2} and \eqref{Dirac2} numerically with the help of Eqs.\eqref{distri2},
\eqref{current} and \eqref{anom}. 
The results are shown in Figs.\ref{fig:Ldis1}, \ref{fig:Tdis} and \ref{fig:current1}. 
The time evolution of longitudinal momentum distributions with fixed transverse momentum is exhibited in Fig.\ref{fig:Ldis1} 
and that of transverse momentum distributions with fixed longitudinal momentum is in Fig.\ref{fig:Tdis}. 
Fig.\ref{fig:current1} shows the time evolution of color current density, electric field strength and quark number density. 
The parameters are set to be $a=\frac{m^2 }{2gE_0 } =0.01$ or $a=0$, and $g=1,\theta =0$. 
To concentrate our attention at first on general features which are common to the Abelian case,
only the distributions of {\lq \lq}blue'' quarks, whose effective coupling to the background is $\frac{1}{2} g$
when $\theta =0$, are presented. 
Hereafter, all figures are shown in the dimension-less unit scaled by $\sqrt{gE_0 } $. 

After the switch-on of the electric field, particles are created with approximately 0 longitudinal momenta. 
Their occupation number is approximately 
$\exp \left( -\pi m_\text{T} ^2 /|w_i |gE_0 \right) $,
which accords with the form expected from the semi-classical tunneling calculation \cite{Casher},
and the transverse momentum distribution exhibits a Gaussian-like form.\footnote{
This is not exactly Gaussian. 
An exact Gaussian distribution is realized in the case of a constant electric field (no back reaction). 
Under a constant electric field, we can calculate the distribution analytically
and can show that the distribution approaches to Gaussian in an asymptotic region where $p_z \to \infty$.} 
In particular, particles in zero mode ($m=0,p_\text{T} =0$) are created as many as possible under the restriction 
of Pauli's exclusion principal $f_{\mathbf{p} ,s} ^i (t) \leq 1$, so that their occupation is exactly equal to one. 
Therefore, the longitudinal momentum distribution with $a=0$ and $p_\text{T} =0$ [Fig.\ref{fig:Ldis1}(b)] 
shows squarish form. 

After created, particles [anti-particles] are accelerated to the direction of 
$w_i g\mathbf{E} $ [$-w_i g\mathbf{E} $]. 
They generate charge current in the direction of $n^a $, which decreases the electric field 
according to the {\lq \lq}Maxwell's" equation \eqref{Maxwell2}.
Then, the direction of the electric field is flipped at some time and the particles start to be decelerated. 
Repeats of this process result in oscillating behavior of the longitudinal momentum distribution, the charge current
and the electric field, which is known as plasma oscillation. 

Other than plasma oscillation, which is a classical dynamics, also the Pauli blocking, which is a quantum effect, 
plays a role in the time evolution of the longitudinal momentum distributions.  
Because of Pauli's exclusion principal, a particle blocks the subsequent pair creation at the point where it locates
in phase space. 
In particular, if a value of a distribution exceeds $1/2$, not only pair creation is suppressed 
but also pair annihilation occurs
(see the second term of the right hand side of Eq.\eqref{initial}). 
Therefore, the distributions get dented when particles cross the line of $p_z =0$ in momentum space.
(Notice that pair creation and annihilation can happen only in the vicinity of the line of $p_z =0$ 
because we now take only a classical gauge field.) 
The effect of the Pauli blocking is the most conspicuous in distributions with $m_\mathrm{T} =0$. 
In this case, because the occupation of the created particles is equal to one, 
particles are totally annihilated when they cross the line of $p_z =0$ [See Fig.\ref{fig:Ldis1}(b)]. 

Collecting the facts above, the momentum distribution function can be approximated by the following equation:
\begin{equation}
f^i _{\mathbf{p} s} (t) \simeq \exp \left( -\frac{\pi m_\text{T} ^2 }{|w_i |gE_0 } \right) 
 \theta \left( p_z (-w_i g\bar{A} _3 -p_z ) \right) . \label{f_approx}
\end{equation}
This equation roughly replicates the distributions obtained by the numerical calculation. 
Using this empirical and analytic expression of the distribution, we analyze the numerical results and 
study the parameter dependence of current and particle number density. 
Equation \eqref{f_approx} is exactly correct for the zero mode with $m_\text{T} =0$. 
In contrast, it loses its accuracy for higher $m_\text{T}$ modes. 
However, it does not matter for our purpose because pair creation of those modes is strongly suppressed
compared with low $m_\text{T}$ modes.  
Although Eq.\eqref{f_approx} is neither an exact one nor obtained by some systematic approximation, 
it reproduces the numerical results with sufficient accuracy for a rough analysis and
is useful because of its simple structure. 

Using Eq.\eqref{f_approx}, the momentum integrations in
the particle number density and the conduction current can be done and we obtain
\begin{align}
N(t) \simeq \frac{4N_f }{(2\pi )^3 } E_0 \sum _{i=1,2,3} (w_i g)^2 
 e^{-\frac{\pi m^2}{|w_i |gE_0 }} \left| \bar{A} _3 (t)\right|  \label{N_approx} \\
j_z ^\text{cond} (t) \simeq -\frac{4N_f }{(2\pi )^3 } E_0 \sum _{i=1,2,3} |w_i g|^3 
 e^{-\frac{\pi m^2}{|w_i |gE_0 }} \bar{A} _3 (t) . \label{j_approx}
\end{align} 
In the derivation of Eq.\eqref{j_approx}, the approximation $p_z /\omega _p \simeq \text{sgn}(p_z) \ (m^2 \ll gE_0 )$ 
has been used, which makes Eq.\eqref{j_approx} overestimate. 
Equation \eqref{j_approx} can explain the plasma oscillation obtained by the numerical calculation. 
Neglecting the polarization current and using Eq.\eqref{j_approx}, 
we can solve the Maxwell equation $\frac{d^2 \bar{A} _3 }{dt^2} = j_z $ and obtain the oscillating electric field
\begin{align}
\bar{A} _3 (t) = -\frac{E_0 }{\Omega } \sin \Omega t \label{A3_approx} \\
E_z (t) = E_0 \cos \Omega t ,
\end{align}
where its frequency is
\begin{align}
\Omega = \sqrt{\frac{4N_f }{(2\pi )^3 } E_0 \sum _i |w_i g|^3 e^{-\frac{\pi m^2}{|w_i |gE_0 }} } . \label{frequency}
\end{align}
Let us introduce the time $t_c$ when the electric field first reduces its strength to zero. 
This $t_c$ gives a typical time scale of variation of the electric field. 
In the present approximation, $t_c $ is given as follows
\begin{align}
t_c = \frac{1}{4} \frac{2\pi }{\Omega } 
 = \frac{\pi }{2} \sqrt{\frac{(2\pi )^3 }{4N_f E_0 \sum _i |w_i g|^3 e^{-\frac{\pi m^2}{|w_i |gE_0 }} } } . \label{t_c}
\end{align}
This gives a smaller value of $t_c $ than the result of the numerical calculations 
because the current \eqref{j_approx} is an overestimate. 
However, their discrepancy is less than factor 2 and Eq.\eqref{t_c} correctly describes the parameter dependence of $t_c $
obtained by the numerical calculations. 
Equation \eqref{t_c} tells us that the stronger the electric field or the larger the coupling constant, 
the shorter time the electric field takes to vanish. 

Substituting Eq.\eqref{A3_approx} into Eq.\eqref{N_approx} leads the oscillating behavior of $N(t)$ 
while the actual $N(t)$ [Fig.\ref{fig:current1}(c)] does not show oscillation.
This discrepancy is because of the incorrectness of the approximation \eqref{f_approx} at $t>t_c $. 
For explanation,
suppose particles having a positive charge $w_i g>0$. 
They get positive momentum at first by acceleration of the electric field, 
and after $t_c$ they go into the negative momentum area in the momentum space because the direction of the field is 
flipped at $t=t_c $. 
The modeled expression \eqref{f_approx} does not describe particles plunging into the negative momentum area. 
If the particles are those in the zero mode $m_\text{T} =0$, they totally disappear when they cross the line of $p_z =0$
due to the Pauli blocking and thus Eq.\eqref{f_approx} is correct at all time. 
If $m_\text{T} \neq 0$, however, they do not totally disappear 
and the simple approximation \eqref{f_approx} becomes incorrect after $t_c$.
That is why Eq.\eqref{N_approx} is not correct at $t>t_c $ and the actual $N(t)$ shows saturation rather than oscillation. 
Under a strong magnetic field, however, $N(t)$ does oscillate in time [Fig.\ref{fig:current_mag2}(c)]
and the present approximation becomes correct
because only the lowest $m_\text{T} $ mode mainly contributes (See Section \ref{sec:mag}). 

\subsection{Polarization current and damping of the electric field} \label{subsec:damping}
The current density \eqref{current} consists of two parts: the conduction current and the polarization current. 
As discussed in the previous subsection, the conduction current is associated with collective motion of particles,
i.e. plasma oscillation. 
In contrast, the polarization current is related with microscopic processes of pair creation. 
It is induced at an instant when pair creation happens. 
In a classical view, when a particle pair is created and becomes on-shell, a distance between them is nonzero 
except the case that their transverse mass is zero. 
Therefore, the pair creation process can be interpreted as creation of an electric dipole,
so that it generates the polarization current. 
(See Ref. \cite{Tanji} for more detailed argument.)

Because of its origin, the polarization current is expected to be induced in the same direction with the electric field. 
In other words, $E_z$ and $j_z ^\text{pol}$ have the same sign. 
Therefore, the polarization current \textit{always} reduces the electric field strength through the Maxwell equation
$\frac{dE_z }{dt} =-j_z$. 
That is, the polarization current causes damping of the electric field. 

\begin{figure}
\begin{center}
\includegraphics[scale=0.55]{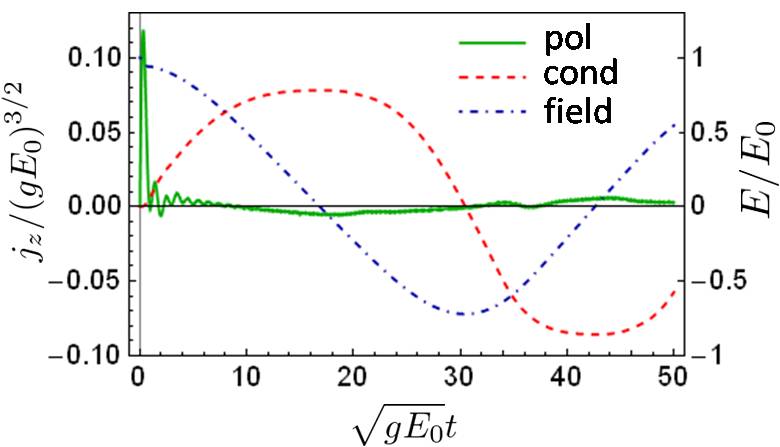} \\ 
\end{center}
 \vskip -\lastskip \vskip -3pt
\caption{Comparison between the polarization and conduction current. $a=0.01 ,g=1, \theta =0$.
         The electric field is also plotted for reference.}
\label{fig:con_pol}
\end{figure}

Nevertheless, the polarization current obtained by our numerical calculation shows irregular behavior 
[Fig.\ref{fig:con_pol}] rather than simple one expected from above argument based on a classical view. 
This is because the system undergoes a complicated evolution due to plasma oscillation and the Pauli blocking,
and furthermore the anomalous distribution, which is included in the integrand of the polarization current \eqref{current},
is sensitive to phases of the Bogoliubov coefficients. 
What is especially remarkable in the plot of the polarization current is a peak 
just after switching the electric field on. 
Because the electric field is turned on suddenly, the vacuum experiences prompt polarization. 
That is why the large polarization current is induced at an early time. 

Although the naive expectation based on a classical view does not hold in this dynamic system, 
damping of electric fields actually happen.  
However, it is slight and hard to be recognized in Fig.\ref{fig:current1}(b)
because the polarization current is far smaller than the conduction current 
(except the time just after the field is switched on) in our parametrization. 
If behavior in longer time is calculated, damping of the field would be observed. 
Indeed, damping behavior can be recognized in the result under a magnetic field which is parallel to 
an electric field [Fig.\ref{fig:current_mag1}(b)], 
because a magnetic field speeds up the time evolution of the system (see Section \ref{sec:mag}).
Besides it, the damping would be quickened by strong coupling. 

Before closing this subsection, let us emphasize that a zero transverse mass mode 
does not contribute to polarization current. 
It is evident from the explicit expression of a polarization current, which contains the factor $m_\text{T}$ 
(the second term of Eq.\eqref{current}). 
In a classical view, the distance between a massless pair is zero when they are created,
so that its electric dipole moment is also zero. 
That is why no polarization current is induced by zero modes.   

\subsection{Interference} \label{subsec:interference}
\begin{figure}
\begin{center}
\includegraphics[scale=0.35]{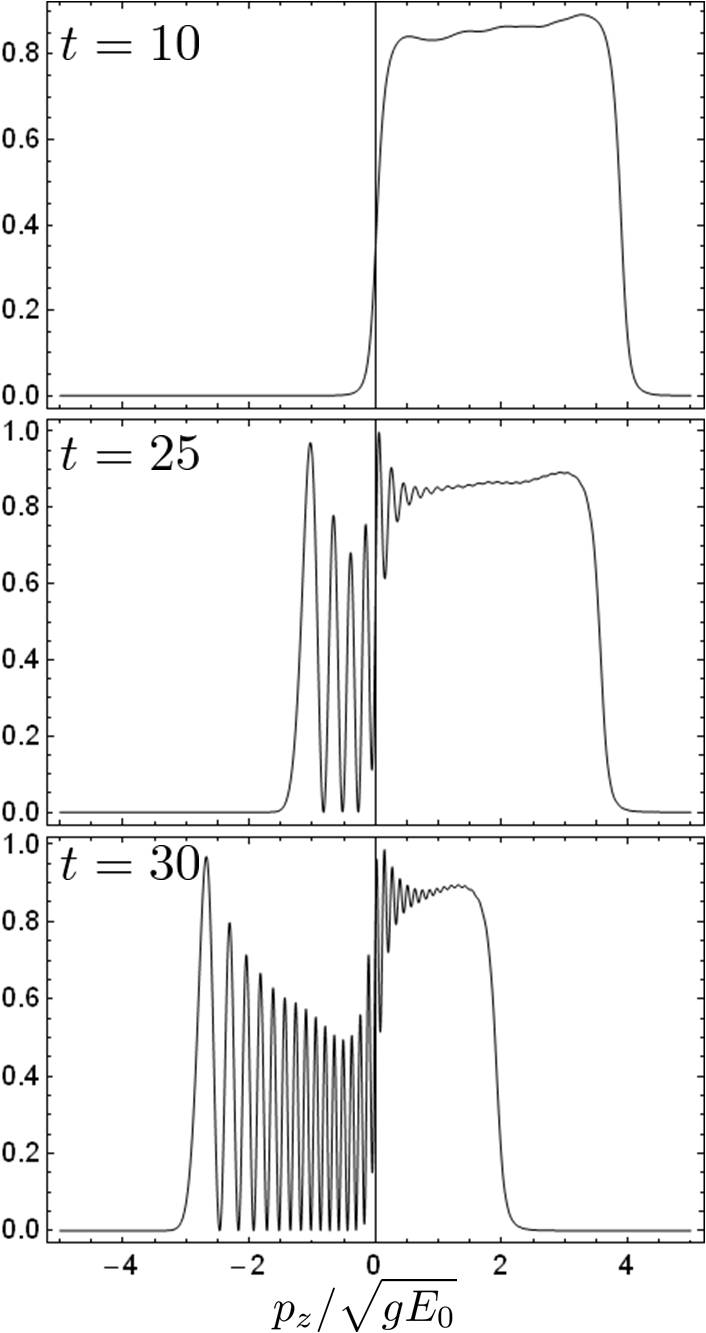} \\ 
\end{center}
 \vskip -\lastskip \vskip -3pt
\caption{Time slices of the longitudinal momentum distribution in Fig.\ref{fig:Ldis1}(a).}
\label{fig:slice}
\end{figure}

One may notice that the longitudinal distribution shows rapid oscillations after $t_c$ [Fig.\ref{fig:Ldis1}].
Fig.\ref{fig:slice} shows time slices of the longitudinal distribution with $a=0.01,g=1$ and $\theta =0$, 
whose whole picture is Fig.\ref{fig:Ldis1}(a). 
Before $t_c$ ($\sim 15/\sqrt{gE_0}$ in this case), the distribution is smooth. 
After $t_c$, particles created before $t_c$ start to cross the line of $p_z =0$. 
Then, the distribution begins to show rapid oscillations. 
These oscillation have been observed also in earlier works where back reaction is taken into account 
\cite{Kluger1991-1993,Kluger1998}. 
These can be interpreted as interference between particles created before $t_c$ and those created after $t_c$. 

Because the problem with back reaction cannot be treated analytically, 
we deal with pair creation under a constant electric field to explain the interference noted above. 
The problem without back reaction can be solved analytically and an explicit expression for distributions is 
available \cite{Tanji}. 
Under a constant electric field, however, particles are accelerated to one direction,
so that there is no event that two distributions overlap in momentum space nor interfere. 
Therefore, we suppose an initial state with a distribution $f_0 (\mathbf{p} )$ instead of a vacuum with no particle 
$|0,\text{in} \rangle $. 
(In this subsection, we omit spin and color indices for simplicity.)
This initial distribution would join a distribution of particles created from an electric field, and they would interfere.  
That initial state is expressed by a two-mode squeezed state as
\begin{align}
|f_0 \rangle = \prod _\mathbf{p} \mathcal{N} _\mathbf{p} ^{-1/2}
 \exp \left[ \frac{(2\pi )^3}{V} F(\mathbf{p} ) a^{\text{in} \, \dagger } _\mathbf{p} 
 b^{\text{in} \, \dagger } _{-\mathbf{p}} \right] |0,\text{in} \rangle , \label{|f0>}
\end{align}
where $\mathcal{N} _\mathbf{p}$ and $F(\mathbf{p} )$ are related with the initial distribution $f_0 (\mathbf{p} )$
as follows
\begin{align}
|F(\mathbf{p} )|^2 = \frac{f_0 (\mathbf{p} )}{1-f_0 (\mathbf{p} )}, \hspace{10pt}
\mathcal{N} _\mathbf{p} = \frac{1}{1-f_0(\mathbf{p} )} 
\end{align}
Notice that a phase of $F(\mathbf{p} )$ is irrelevant to $f_0 (\mathbf{p} )$.
Because now $f_0 (\mathbf{p} )$ is given by hand, there is no criterion to decide a phase of $F(\mathbf{p} )$.
However, if $f_0 (\mathbf{p} )$ is a distribution of particles created from an electric field, 
a phase of $F(\mathbf{p} )$ is automatically determined by its time history of evolution. 

Letting this system evolve under the electric field which is given by the gauge 
$A^\mu =\left( 0,\mathbf{A} (t) \right)$, we obtain the distribution function at time $t$:
\begin{align}
\tilde{f} _{\mathbf{p}} (t) 
 &= \langle f_0 |a^{\dagger } _{\mathbf{p}} (t)a_{\mathbf{p}} (t) |f_0 \rangle \frac{(2\pi )^3 }{V} \notag \\
 &= f_0 \left( \mathbf{p} +w_i g \mathbf{A} (t) \right) 
   +\left \{ 1-2f_0 \left( \mathbf{p} +w_i g \mathbf{A} (t) \right) \right \} f_{\mathbf{p} } (t)  \label{initial} \\
 &\hspace{10pt}
   +2\left \{ 1-f_0 \left( \mathbf{p} +w_i g \mathbf{A} (t) \right) \right \} 
    \text{Re} \left[ \alpha _{\mathbf{p} } (t)\beta ^{*} _{\mathbf{p} } (t) F(\mathbf{p} +w_i g \mathbf{A} (t) )\right] . \notag
\end{align}
The first term of the right hand side represents the initial particles accelerated by the field.
$f_{\mathbf{p}} (t) $ in the second term is the distribution of particles created from the field,
which is defined by Eq.\eqref{distri}. 
The factor $\left \{ 1-2f_0 \left( \mathbf{p} +w_i g \mathbf{A} (t) \right) \right \}$ expresses the effect of
the Pauli blocking: initial particles suppress the subsequent pair creation, and 
pair annihilation occurs if the initial occupation exceeds $1/2$.
The third term describes interference between the initial particles and those created from 
the field.\footnote{This interference term has been missed in the previous paper \cite{Tanji}. } 
Now $\alpha _{\mathbf{p}} (t)$ and $\beta _{\mathbf{p}} (t)$ are the Bogoliubov coefficients giving
the distribution $f_{\mathbf{p}} (t)$. 
Although the distributions $f_{\mathbf{p}} (t)$ and $f_0 (\mathbf{p} )$ are independent of 
the phases of the Bogoliubov coefficients or $F(\mathbf{p} )$, the third term is sensitive to those phases. 
Because of this term, the distribution shows rapid oscillations when the two distributions overlap in momentum space. 

\begin{figure}
 \begin{tabular}{cc}
  \begin{minipage}{0.5\textwidth}
   \begin{center}
    \includegraphics[scale=0.5]{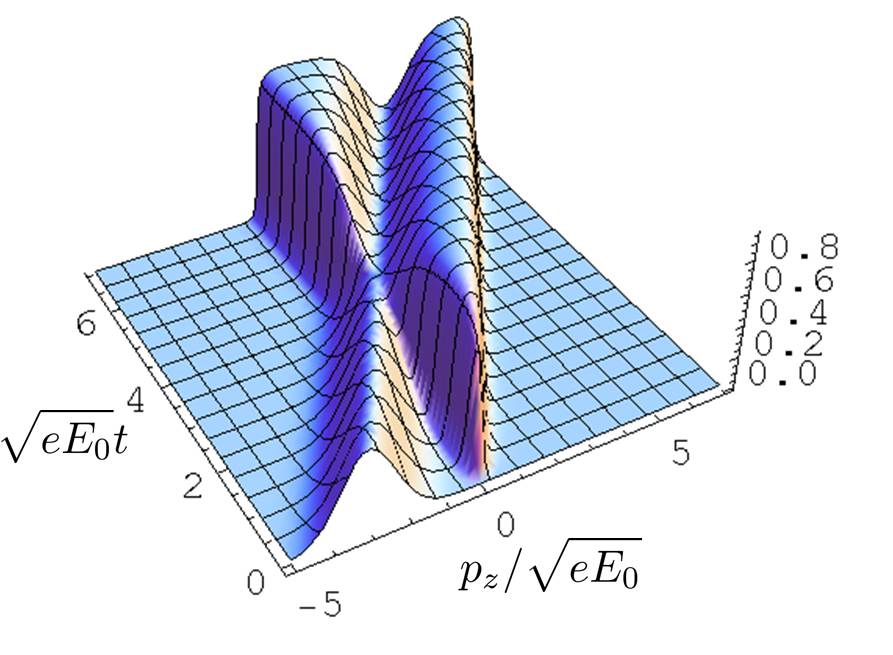} 
   \end{center}
  \end{minipage} & 
  \begin{minipage}{0.5\textwidth}
   \begin{center}
    \includegraphics[scale=0.5]{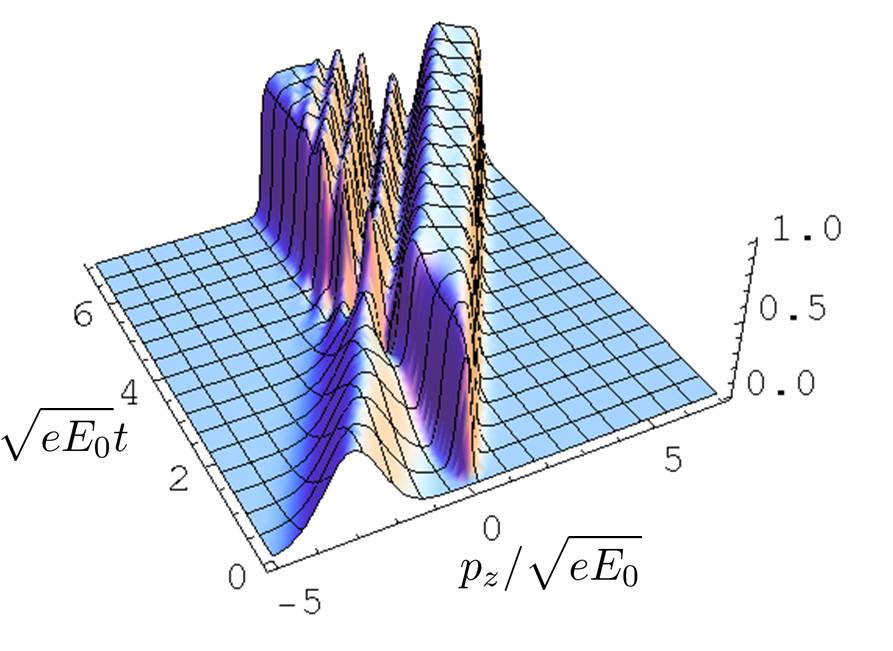} 
   \end{center}
  \end{minipage} \\
  (a) neglecting the interference term &
  (b) $F(\mathbf{p} ) =  \sqrt{\frac{f_0 (\mathbf{p} )}{1-f_0 (\mathbf{p} )}} $ \\
  \begin{minipage}{0.5\textwidth}
   \begin{center}
    \includegraphics[scale=0.5]{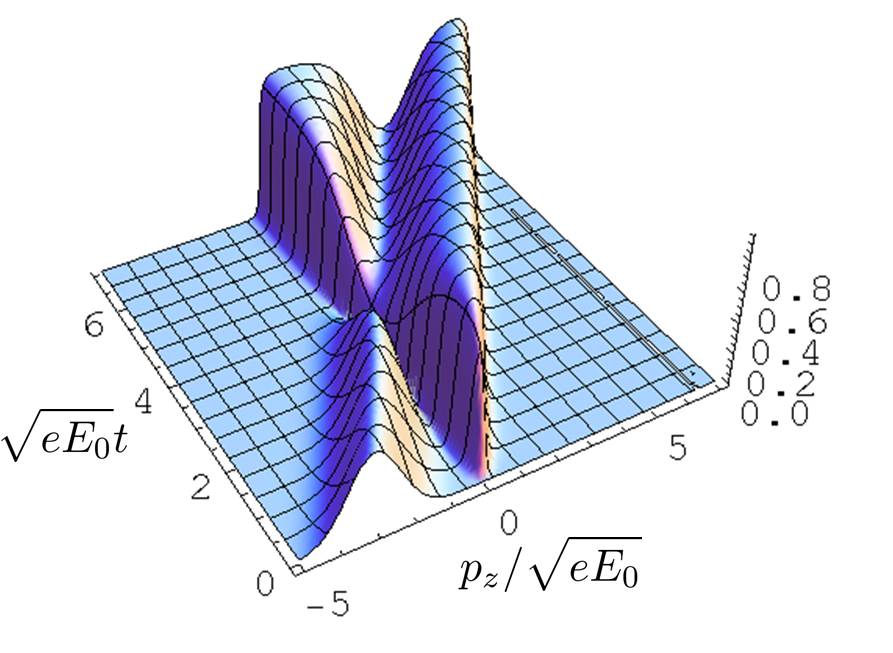} 
   \end{center}
  \end{minipage} & 
  \begin{minipage}{0.5\textwidth}
   \begin{center}
    \includegraphics[scale=0.5]{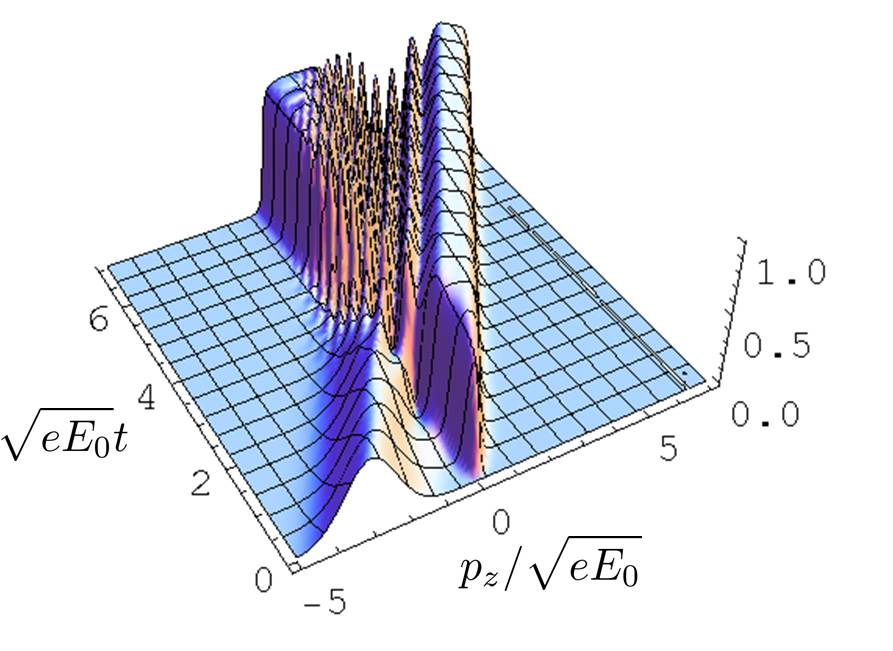} 
   \end{center}
  \end{minipage} \\
  (c) $F(\mathbf{p} ) =  \sqrt{\frac{f_0 (\mathbf{p} )}{1-f_0 (\mathbf{p} )}} e^{ip_z ^2/eE_0} $ &
  (d) $F(\mathbf{p} ) =  \sqrt{\frac{f_0 (\mathbf{p} )}{1-f_0 (\mathbf{p} )}} e^{-ip_z ^2/eE_0} $
 \end{tabular}
 \caption{Distribution functions under a constant electric field with the initial distribution 
  $f_0 (\mathbf{p})= \frac{1}{2} e^{-(p/\sqrt{eE_0 } +5)^2}$. $a=0.01, e=1,p_\text{T} =0$. }
 \label{fig:initial}
\end{figure}

As an illustration, we plot in Fig.\ref{fig:initial} the distribution \eqref{initial} with the initial distribution 
$f_0 (\mathbf{p} )= \frac{1}{2} e^{-(p/\sqrt{eE_0 } +5)^2}$ 
under a constant electric field which is switched on at $t=0$. 
For simplicity, results in Abelian theory, which is obtained by replacing $w_i g$ with $e$ in \eqref{initial}, are shown. 
The parameters are chosen as $a=0.01, e=1$. 
For reference, we show in Fig.\ref{fig:initial}(a) the result in which the interference term is neglected. 
We can see clearly the effect of the Pauli blocking. 
In the presence of the third term, the distributions show oscillations 
of which contour is along classical trajectories of particles [Fig.\ref{fig:initial}(b),(d)]. 
What is remarkable is Fig.\ref{fig:initial}(c), in which a phase factor of $F(\mathbf{p} )$ is set to be $e^{ip_z ^2 /eE_0 }$. 
In this case, oscillations seen in Fig.\ref{fig:initial}(b) disappears. 
That is to say, the factor $e^{ip_z ^2 /eE_0 }$ cancels the phases of $\alpha _{\mathbf{p} }$ and $\beta ^* _{\mathbf{p} }$.
This is reasonable because the phase factors of $\alpha _{\mathbf{p} }$ and $\beta _{\mathbf{p} }$ are expected to be
$e^{iS_{cl} }$ and $e^{-iS_{cl} }$, respectively, in which $S_{cl}$ is the classical action of a particle 
under the constant electric field: 
\begin{align}
S_{cl} \approx -\frac{1}{2} \frac{p_z ^2}{eE_0 } \hspace{20pt} (p_z \to \infty ).
\end{align}
Hence, if one want to reproduce the situation in which a distribution of particles created from the electric field
interfere with a distribution of particles which are \textit{also} created from the electric field, 
it is natural to assume $F(\mathbf{p} )$ has the same phase factor with $\alpha _{\mathbf{p} } \beta ^* _{\mathbf{p} }$,
i.e. $e^{-ip_z ^2 /eE_0 }$. 
A distribution in such case is plotted in Fig.\ref{fig:initial}(d). 
Rapid oscillations similar to those seen in the distribution with back reaction are obtained. 

Note that like as the polarization current, the interference term has no contribution from a zero transverse mass mode. 
That is because the occupation $n_{\mathbf{p} ,s} (t) =|\beta _{\mathbf{p} s} (t)|^2$ of a zero mode takes the maximum value 1,
and there is the constraint $|\alpha _{\mathbf{p} s} |^2 +|\beta _{\mathbf{p} s} |^2 =1$,
so that $\alpha _{\mathbf{p} } (t)\beta ^{*} _{\mathbf{p} } (t)$ is 0 for this mode. 

These rapid oscillations are a key ingredient for seeming irreversibility of time evolution \cite{Habib}. 
Because of this interference as well as a polarization current, effective dissipation of energy from the electric field
to the quantum fields happens. 
For example, the number density in Fig.\ref{fig:current1}(c) shows saturation, 
and there is seemingly an arrow of time, although the equations of motion \eqref{Maxwell1} and \eqref{Dirac1} hold 
time-reversal symmetry. 
In contrast, if there is no polarization current nor the interference term, effective dissipation would not happen
and evolution of the system would be periodic in time. 
We shall demonstrate it by studying pair creation under a strong magnetic field in Section \ref{subsec:enhance}
(see Fig.\ref{fig:current_mag2}(c)). 

\begin{figure}[t]
 \begin{tabular}{lccc}
  \begin{minipage}{0.12\textwidth}
   \begin{center}
    \includegraphics[scale=0.5]{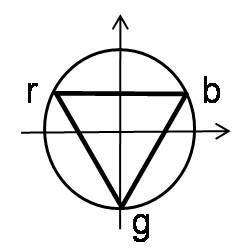} \\ 
    $\theta =0$
   \end{center}
  \end{minipage} & 
  \begin{minipage}{0.28\textwidth}
   \begin{center}
    \includegraphics[scale=0.35]{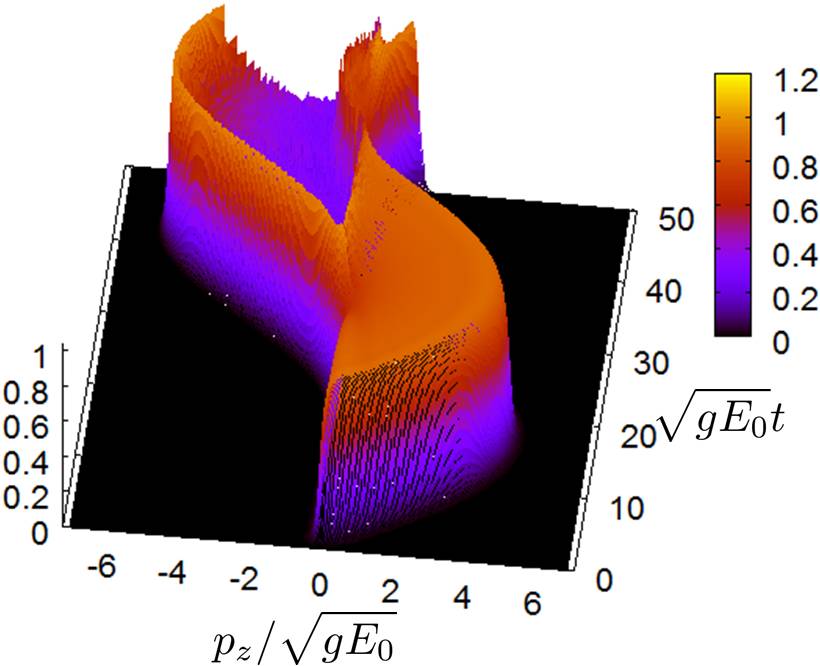} 
   \end{center}
  \end{minipage} & 
  \begin{minipage}{0.28\textwidth}
   \begin{center}
    \includegraphics[scale=0.35]{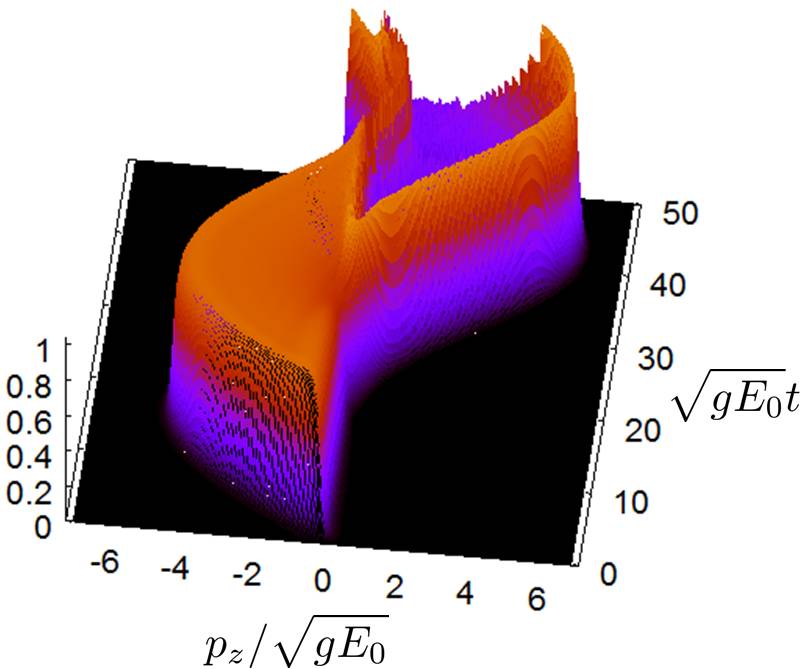} 
   \end{center}
  \end{minipage} & 
  \begin{minipage}{0.28\textwidth}
   \begin{center}
    \text{{\Huge 0 }}
   \end{center}
  \end{minipage} \\
  \begin{minipage}{0.12\textwidth}
   \begin{center}
    \includegraphics[scale=0.5]{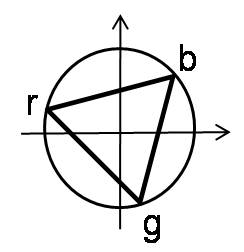} \\ 
    $\theta =\pi/12 $
   \end{center}
  \end{minipage} & 
  \begin{minipage}{0.28\textwidth}
   \begin{center}
    \includegraphics[scale=0.35]{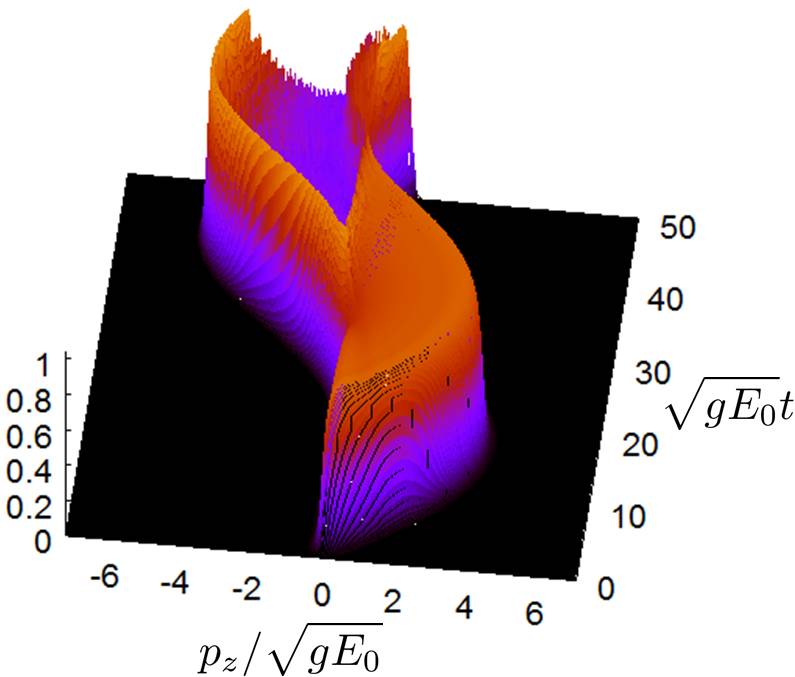} 
   \end{center}
  \end{minipage} & 
  \begin{minipage}{0.28\textwidth}
   \begin{center}
    \includegraphics[scale=0.35]{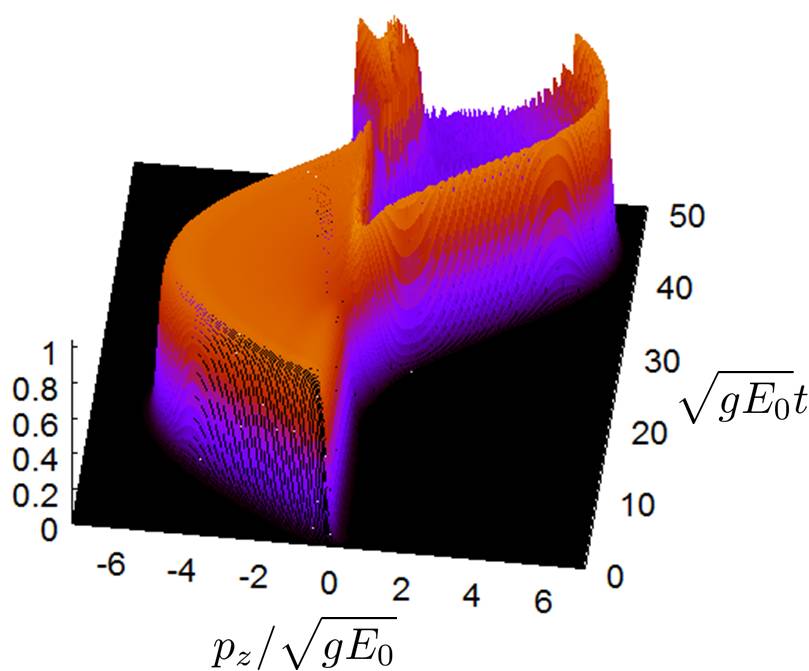} 
   \end{center}
  \end{minipage} & 
  \begin{minipage}{0.28\textwidth}
   \begin{center}
    \includegraphics[scale=0.35]{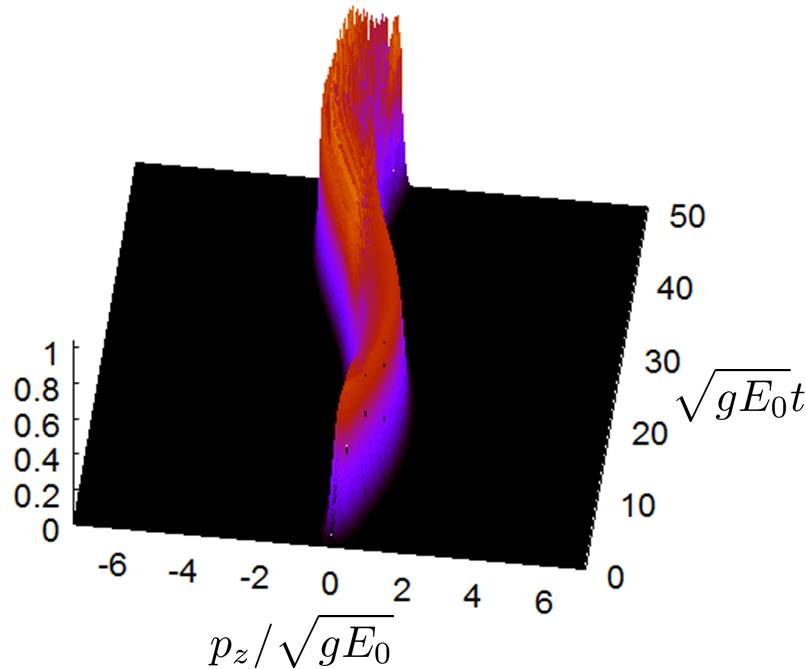} 
   \end{center}
  \end{minipage} \\
  \begin{minipage}{0.12\textwidth}
   \begin{center}
    \includegraphics[scale=0.5]{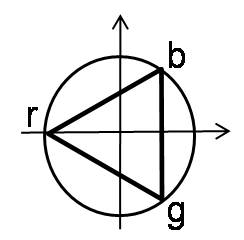} \\ 
    $\theta =\pi/6 $
   \end{center}
  \end{minipage} & 
  \begin{minipage}{0.28\textwidth}
   \begin{center}
    \includegraphics[scale=0.35]{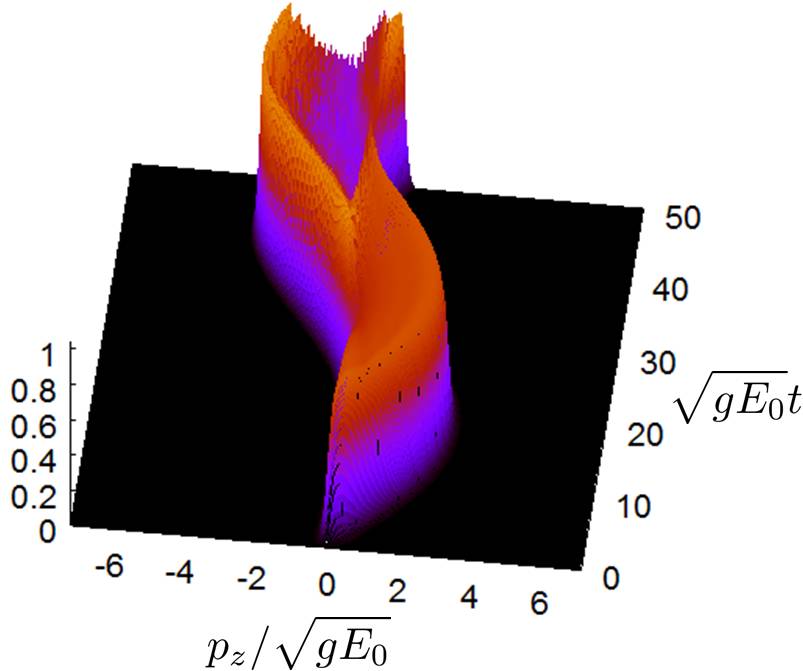} 
   \end{center}
  \end{minipage} & 
  \begin{minipage}{0.28\textwidth}
   \begin{center}
    \includegraphics[scale=0.35]{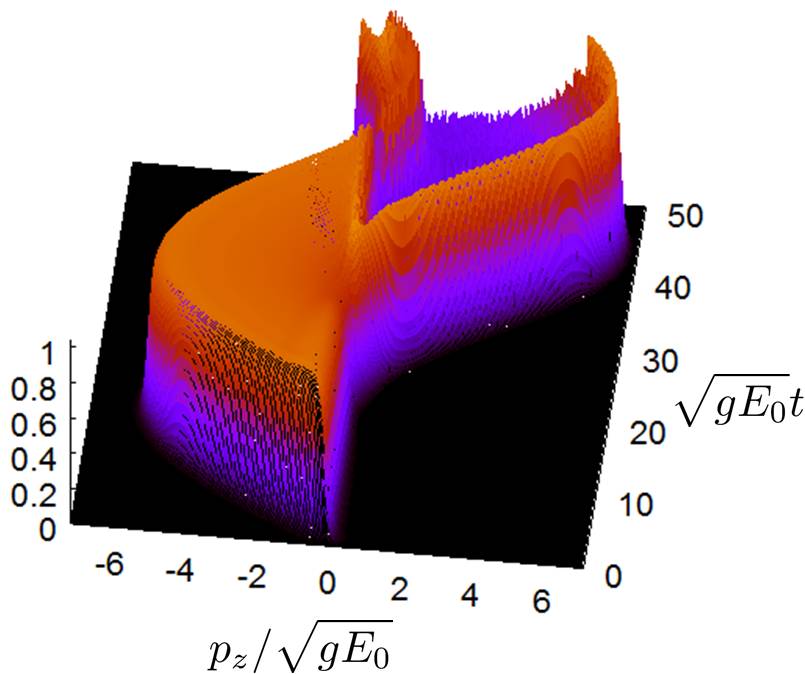} 
   \end{center}
  \end{minipage} & 
  \begin{minipage}{0.28\textwidth}
   \begin{center}
    \includegraphics[scale=0.35]{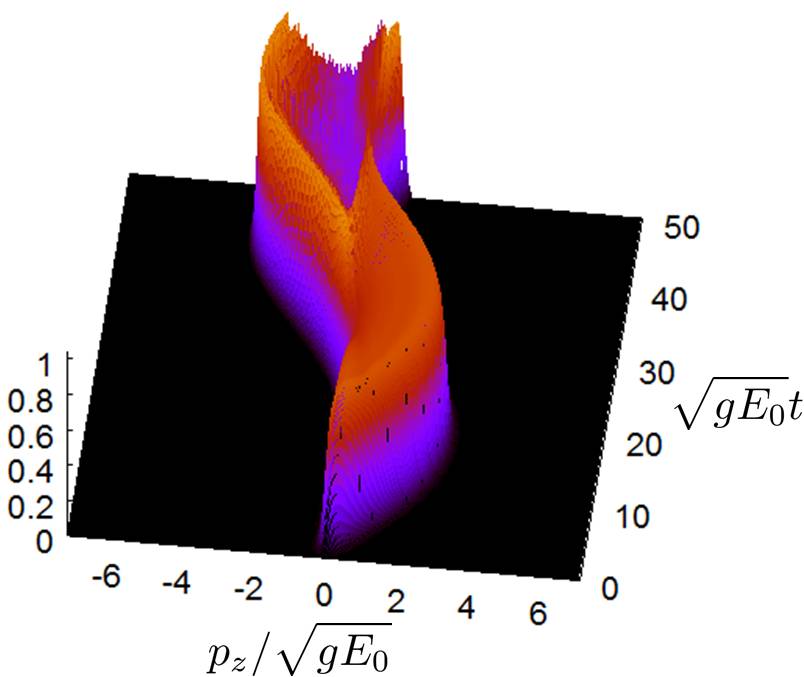} 
   \end{center}
  \end{minipage} \\[+2cm]
  {} & {\lq\lq}blue" & {\lq\lq}red" & {\lq\lq}green" \\
 \end{tabular}
 \caption{Color direction dependence of the longitudinal momentum distributions. 
          $a=0.01,g=1,p_\text{T} =0$.}
 \label{fig:Ldis2}
\end{figure}

\subsection{Color direction dependence} \label{subsec:color}
As shown in Section \ref{sec:frame}, direction of a color electric field in color space can be characterized by 
the Casimir invariant $C_2$ or the angle $\theta$ in a gauge invariant way. 
Therefore, physical values may generally depend on $C_2$ or $\theta$. 
Casimir dependence of transverse distribution\footnote{Their definition of transverse distribution is different from us. 
It actually must be called pair creation probability rather than transverse distribution. } 
under a constant color electric field has been studied in Ref. \cite{Cooper2008}. 
However, as far as the author knows, 
color direction dependence incorporated with back reaction has not been investigated.  

Fig.\ref{fig:Ldis2} shows the color direction dependence of the longitudinal momentum distributions
($a=0.01,g=1,p_\text{T} =0$). 
We can understand the behavior of these distributions based on the values of the effective coupling $w_i g$. 
For example, when $\theta =0$, the distributions of {\lq\lq}blue" and {\lq\lq}red" quarks are symmetric 
in momentum space because $w_1 =-w_2$, 
and {\lq\lq}green" quarks are not at all created since $w_3 =0$. 
Concerning the momentum distributions, the result greatly depends on the color direction of the electric field. 
This is a matter of course because the distributions of unconfined quarks are not color singlet object. 

\begin{figure}
 \begin{tabular}{ccc}
  \begin{minipage}{0.33\textwidth}
   \begin{center}
    \includegraphics[scale=0.42]{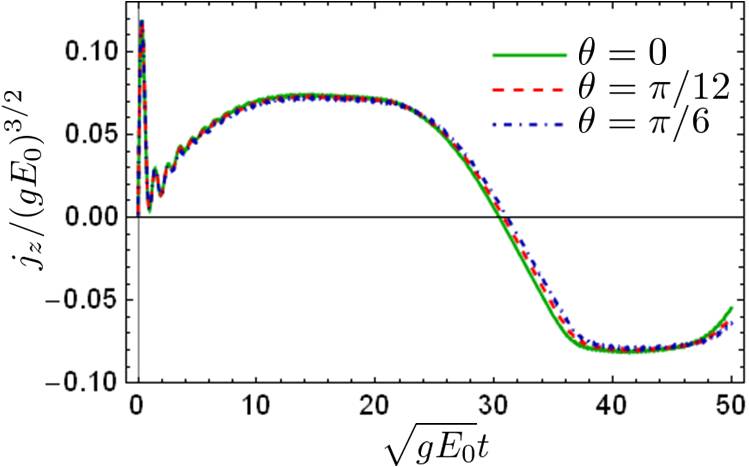} 
   \end{center}
  \end{minipage} & \hspace{-0.5cm}
  \begin{minipage}{0.33\textwidth}
   \begin{center}
    \includegraphics[scale=0.4]{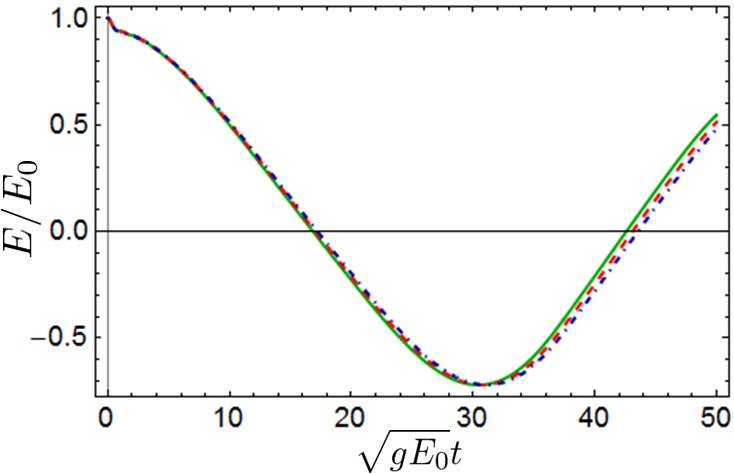} 
   \end{center}
  \end{minipage} & \hspace{-0.5cm}
  \begin{minipage}{0.33\textwidth}
   \begin{center}
    \includegraphics[scale=0.4]{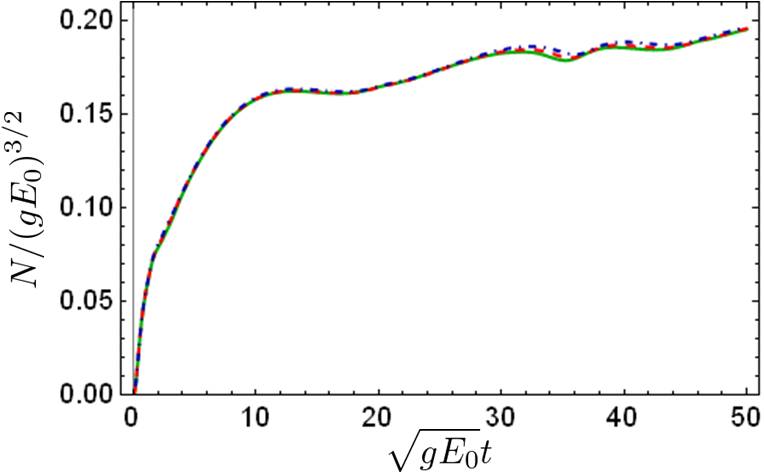} 
   \end{center}
  \end{minipage} \\
  (a) color current density &
  (b) electric field &
  (c) quark number density
 \end{tabular}
 \caption{Color direction dependence of the current density, the electric field and the quark number density. 
          $a=0.01,g=1$.}
 \label{fig:current2}
\end{figure}

Color direction dependence of the charge current density, the field strength and the quark number density is 
shown in Fig.\ref{fig:current2}. 
Unlike the momentum distributions, the color direction dependence is very small for these quantities. 
This is because the current and the particle number are obtained by summing up the colors, 
so that they depend on $\theta $ only through the elementary symmetric polynomials of $w_i$. 
The relations between $\theta $ and the elementary symmetric polynomials of $w_i $ are
\begin{gather}
w_1 +w_2 +w_3 = 0 \label{w^1} \\
w_1 ^2 +w_2 ^2 +w_3 ^2 = \frac{1}{2} \label{w^2} \\
w_1 w_2 w_3 = -\frac{1}{12\sqrt{3} } \sin 3\theta . \label{w^3}
\end{gather}
The first equation is the trace of Eq.\eqref{UnTU}, 
the second is the trace of the square of Eq.\eqref{UnTU} and the third is the determinant of Eq.\eqref{UnTU}. 
Because $w_1 +w_2 +w_3$ and $w_1 ^2 +w_2 ^2 +w_3 ^2$ are independent of $\theta$ and only $w_1 w_2 w_3$ has the dependence 
on $\theta$, the field quantities depend on the color direction parameter $\theta$ very weakly. 

As an illustration, see Eqs.\eqref{N_approx} and \eqref{j_approx}. 
Because our attention is now on the regime where pair creation strongly happens, i.e. $m^2 /gE_0 \ll 1$,
the factor $\exp \left( -\frac{\pi m^2}{|w_i |gE_0 } \right)$ is nearly equals to 1 
and its $\theta $-dependence is negligible.  
Therefore, leading dependence on $\theta$ comes from the factor $(w_i g)^2$ or $|w_i g|^3$. 
However, the factor $(w_i g)^2$ brings no $\theta $-dependence because of Eq.\eqref{w^2}. 
That is why color direction dependence of the number density is very small.  
In the case of the current density, although the factor $\sum _i |w_i g|^3$ does bring $\theta $-dependence,
one can show that the $\theta $-dependence is numerically small by substituting Eq.\eqref{weight} 
explicitly into $\sum _i |w_i g|^3$. 
It varies between $0.24g^3$ and $0.25g^3$.  

\subsection{Pressure} \label{subsec:pressure}
The initial state with the longitudinal electric field $\mathbf{E} ^a =(0,0,E)n^a $ is quite anisotropic: 
longitudinal pressure is $P_\text{L} =-E^2 /2$ and transverse pressure is $P_\text{T} =E^2 /2$. 
In contrast, pressure is locally isotropic in a locally thermalized quark-gluon plasma phase. 
Hence, if longitudinal color electric fields are formed in the initial stage of heavy-ion collisions 
and if local thermalization is realized in the later stage, 
isotropization of pressure must be achieved during the evolution of the system. 
In this subsection, we examine 
how the initial anisotropic pressure evolves under the influences of pair creation and its back reaction. 

\begin{figure}
 \begin{center}
  \includegraphics[scale=0.42]{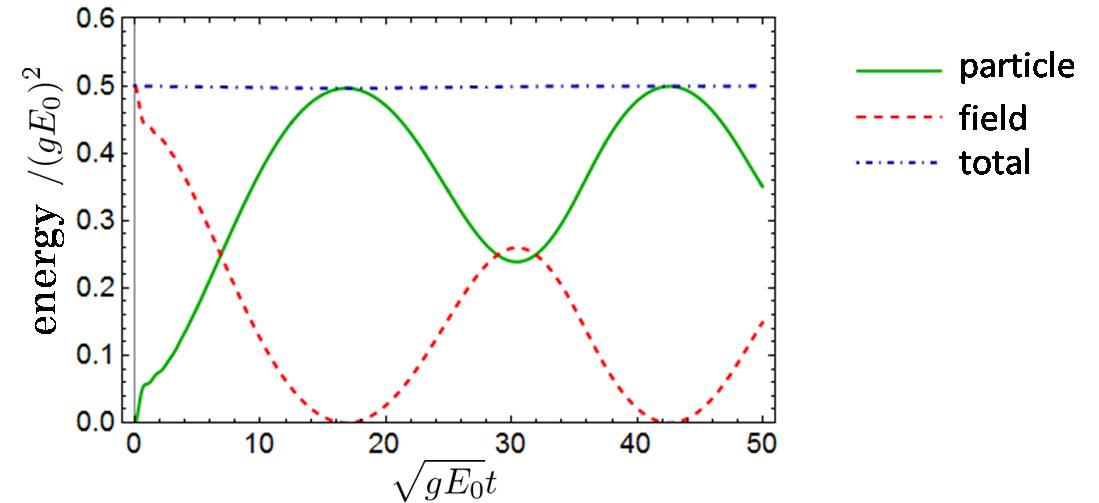} \\ 
 \end{center}
 \vskip -\lastskip \vskip -6pt
 \caption{Energy balance.}
 \label{fig:energy}
\end{figure}

Pressure generated by quarks is given by the expectation of the symmetric energy-momentum tensor: 
\begin{equation}
\begin{split}
\langle 0,\text{in} |:\Theta ^{\mu \nu} :|0,\text{in} \rangle
 &= -\frac{i}{4} \langle 0,\text{in} |:\bar{\psi } \left( \gamma ^\mu \overleftrightarrow{\partial} ^\nu
                      +\gamma ^\nu \overleftrightarrow{\partial} ^\mu \right) \psi :|0,\text{in} \rangle \\
 &= \text{diag} \left( \mathcal{E} ,P_\text{T} ,P_\text{T} ,P_\text{L} \right) , \label{energy-momentum}
\end{split}
\end{equation}
which is regularized by the normal ordering. 
Notice that $\langle \Theta ^{\mu \nu} \rangle $ is diagonal because we are in the center of mass frame.
Inserting Eq.\eqref{psi} into Eq.\eqref{energy-momentum} and using Eq.\eqref{Bogo},
one can express $\mathcal{E} $ (energy density; Fig.\ref{fig:energy}), $P_\text{L} $ and $P_\text{T} $ in terms of 
the momentum distribution $f^i _{\mathbf{p} s} (t) $ and the anomalous distribution $g^i _{\mathbf{p} s} (t) $: 
\begin{gather}
\mathcal{E} = 2N_f \sum _{i=1,2,3} \sum _{s=\uparrow ,\downarrow} 
              \int \! \frac{d^3 p}{(2\pi )^3 } \omega _p f^i _{\mathbf{p} s} (t) \\
P_\text{L} = 2N_f \sum _{i=1,2,3} \sum _{s=\uparrow ,\downarrow}  \int \! \frac{d^3 p}{(2\pi )^3 } 
             \left[ \frac{p_\text{L} ^2 }{\omega _p } f^i _{\mathbf{p} s} (t)
             +\frac{p_\text{L} m_\text{T} }{\omega _p } g^i _{\mathbf{p} s} (t) \right] \ \label{PL} \\ 
P_\text{T} = 2N_f \sum _{i=1,2,3} \sum _{s=\uparrow ,\downarrow}  \int \! \frac{d^3 p}{(2\pi )^3 } 
             \frac{1}{2} \left[ \frac{p_\text{T} ^2 }{\omega _p } f^i _{\mathbf{p} s} (t)
             -\frac{p_\text{L} p_\text{T} ^2 }{m_\text{T} \omega _p } g^i _{\mathbf{p} s} (t) \right] . \label{PT}
\end{gather}
The terms containing $f^i _{\mathbf{p} s} (t)$ are contributions from real particles, 
which are common to classical theory. 
What is peculiar to our quantum field theoretical calculation is the presence of the terms including $g^i _{\mathbf{p} s} (t)$, 
which exhibit pressure generated by pair creation processes. 

One can confirm that the relation
\begin{equation}
\begin{split}
\langle 0,\text{in} |:\Theta ^\mu _{\ \mu} :|0,\text{in} \rangle 
 &= \mathcal{E} -2P_\text{T} -P_\text{L} \\
 &= 2N_f \sum _{i=1,2,3} \sum _{s=\uparrow ,\downarrow} \int \! \frac{d^3 p}{(2\pi )^3 } 
    \left[ \frac{m^2 }{\omega _p } f^i _{\mathbf{p} s} (t) 
    -\frac{p_\text{L} m^2 }{\omega _p m_\text{T} } g^i _{\mathbf{p} s} (t) \right] \\
 &= m\langle 0,\text{in} |:\bar{\psi} \psi :|0,\text{in} \rangle
\end{split}
\end{equation}
holds.\footnote{Trace anomaly is not included in our treatment because we do not consider quantum gauge fields.} 

\begin{figure}
 \begin{tabular}{cc}
  \begin{minipage}{0.5\textwidth}
   \begin{center}
    \includegraphics[scale=0.45]{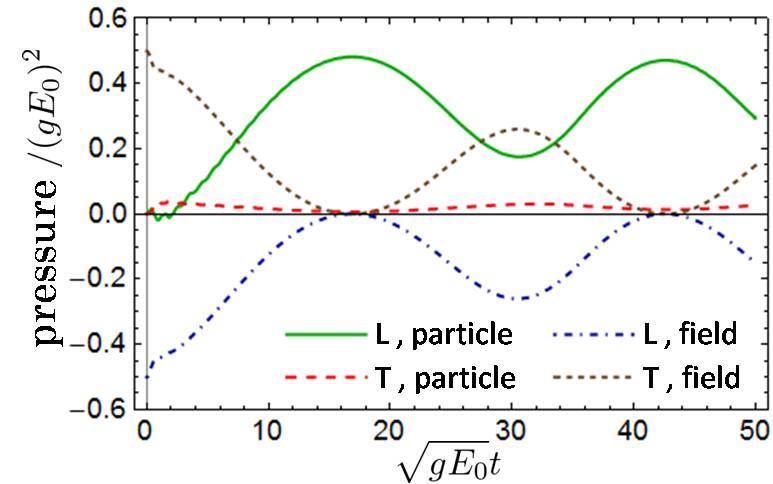} 
   \end{center}
  \end{minipage} &
  \begin{minipage}{0.5\textwidth}
   \begin{center}
    \includegraphics[scale=0.45]{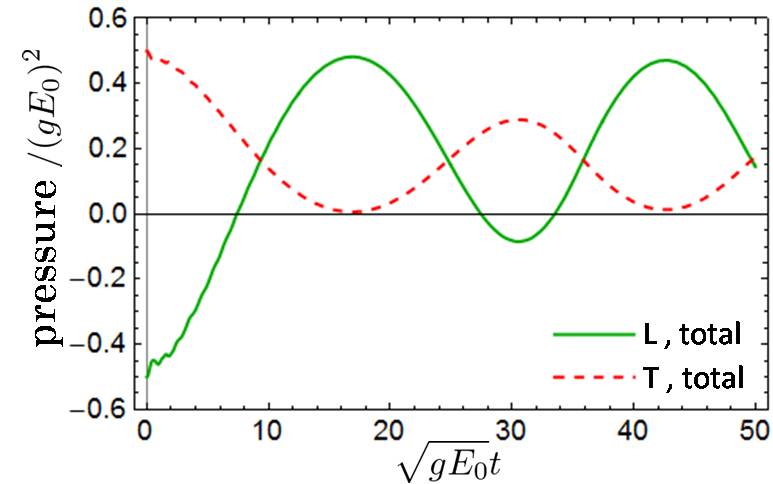} 
   \end{center}
  \end{minipage} \\
  (a) pressure generated by particles and the field &
  (b) total pressure
 \end{tabular}
 \caption{Time evolution of pressure. $a=0.01,g=1,\theta =0$.
          L and T denote longitudinal and transverse, respectively. }
 \label{fig:pressure}
\end{figure}

The result of our numerical calculation is shown in Fig.\ref{fig:pressure}. 
The parameters are set to be $a=0, g=1$ and $\theta =0$. 
Due to pair creation and subsequent acceleration by the electric field, 
the particles generate positive pressure in the longitudinal direction,
whereas transverse pressure by the particles is relatively small 
because the particles are accelerated only to the longitudinal direction. 
As pressure by the particles increases,
pressure of the electric field is weakened owing to back reaction. 
The negative longitudinal pressure at the initial time is compensated by pressure generated by the particles. 
Although full isotropization is not achieved because this system is collision-less plasma, 
degree of anisotropy is moderated due to pair creation. 

\section{Effects of magnetic fields} \label{sec:mag} 
In the color glass condensate framework, it has been shown that longitudinal color magnetic fields are produced
in addition to color electric fields just after heavy-ion collisions \cite{Kovner,Lappi}. 
In this section, we study the effects of a longitudinal color magnetic field on pair creation. 
Let us consider the situation such that both electric and magnetic fields are along the $z$-axis 
in configuration space and along $n^a$ in color space.
This setup realizes a gauge field configuration with nonzero topological charge 
$F_{\mu \nu} ^a \tilde{F} ^{a \, \mu \nu} \neq 0$ which is predicted by the color glass condensate. 

Because this system is spatially uniform, induced current is also uniform, 
so that the magnetic field is not changed by back reaction. 
Therefore, the magnetic field acts as only a catalyst in this situation. 
To treat back reaction to a magnetic field, a non-uniform field must be considered. 
However, it is beyond the scope of our present paper. 

\subsection{Enhancement of pair creation} \label{subsec:enhance}
Under a longitudinal magnetic field, pair creation is enhanced,
so that the time evolution of the system becomes faster due to the enhanced back reaction \cite{Tanji}. 
This phenomenon is caused by appearance of Landau levels and spin-magnetic field interaction. 

Due to interaction with the magnetic field, transverse momentum of created particles is discretized into Landau levels, and
degeneracy between modes with spin parallel and antiparallel to the magnetic field, which are denoted by 
$\uparrow$ and $\downarrow$ each in the following, is broken:
\begin{equation}
p_\text{T} ^2 \longrightarrow 2n_s |w_i | gB, \label{LL1}
\end{equation}
where
\begin{equation}
n_s = \left \{ \begin{array}{lll} 
n & &\text{for $s=\uparrow$\ } \\
n+1 & \raisebox{0.5\normalbaselineskip}[0pt][0pt]{$(n=0,1,\cdots )$} & \text{for $s=\downarrow$.}
\end{array} \right.
\end{equation}
The expressions of field quantities under the longitudinal magnetic field are available from those under 
no magnetic field by the replacements \eqref{LL1} and
\begin{equation}
\begin{split}
m_\text{T} ^2 &\longrightarrow m_{\text{T},s} ^2 = m^2 +2n_s |w_i | gB \\
\omega _p &\longrightarrow \omega _{\mathbf{p} ,s} = \sqrt{p_z ^2 +m_{\text{T},s} ^2 } \\
\int \! \frac{d^2 p_\text{T} }{(2\pi )^2 } &\longrightarrow \frac{|w_i |gB}{2\pi } \sum _{n=0 } ^{\infty} \ ,
\end{split} \label{LL2}
\end{equation}
where $\mathbf{p} $ in $\omega _{\mathbf{p} ,s} $ is the abbreviated expression for $(p_z ,n)$. 
Notice that the transverse mass with $n_s =0$ is independent of $B$, while those with higher $n_s $ depend on $B$. 
Therefore, a strong magnetic field makes particles in higher modes \lq\lq heavy" and suppresses their pair creation.
In contrast, creation of $(n_s =0)$-particles is not at all suppressed. 

Not only the creation of the lowest mode is not suppressed, the magnetic field enhances field quantities such as current and 
total particle number.
That is because the number of modes degenerating in a unit transverse area is proportional to $B$.

For example, the charge current \eqref{current} is replaced by
\begin{equation}
j_z (t) 
 = 2N_f \sum _{i=1,2,3} \sum _{s=\uparrow ,\downarrow } w_i g \frac{|w_i |gB}{2\pi } \sum _n \int \! \frac{dp_z }{2\pi } 
   \left[ \frac{p_z }{\omega _{\mathbf{p} ,s} } f^i _{\mathbf{p} s} (t) 
   +\frac{m_{\mathrm{T} ,s} }{\omega _{\mathbf{p} ,s} } g^i _{\mathbf{p} s} (t) \right] .
\end{equation}
In a strong magnetic field $gB\gtrsim gE\gg m^2$, this may be approximated as follows
\begin{equation}
j_z (t) \simeq 
 2N_f \sum _{i=1,2,3} w_i g \frac{|w_i |gB}{2\pi } \int \! \frac{dp_z }{2\pi } 
 \left[ \frac{p_z }{\omega _{p_z ,n=0 ,\uparrow} } f^i _{p_z ,n=0 ,\uparrow} (t)
 +\frac{m}{\omega _{p_z ,n=0 ,\uparrow} } g^i _{p_z ,n=0 ,\uparrow} (t) \right] ,
\end{equation}
which is proportional to $B$. 

\begin{figure}
 \begin{tabular}{ccc}
  \begin{minipage}{0.33\textwidth}
   \begin{center}
    \includegraphics[scale=0.42]{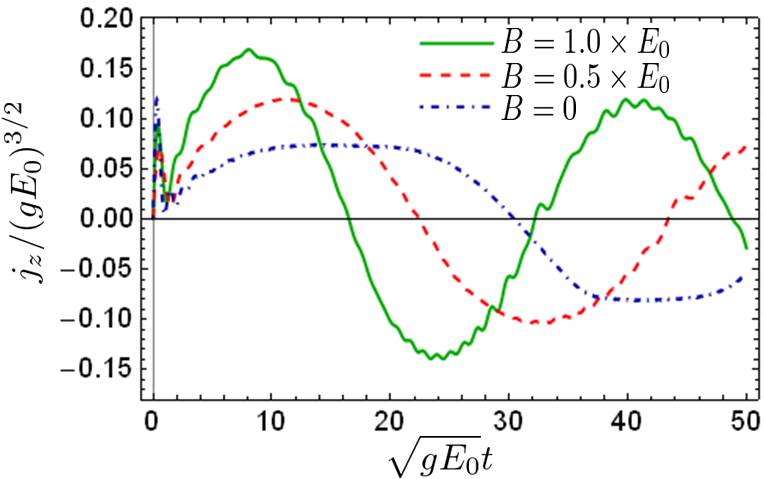} 
   \end{center}
  \end{minipage} & 
  \begin{minipage}{0.33\textwidth}
   \begin{center}
    \includegraphics[scale=0.4]{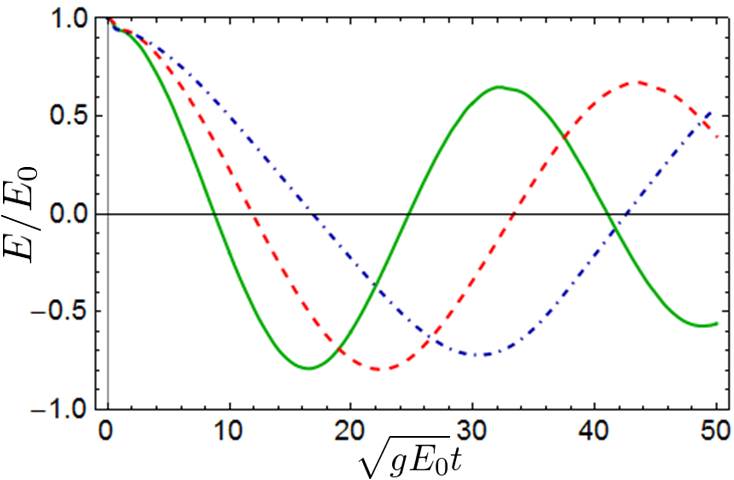} 
   \end{center}
  \end{minipage} &
  \begin{minipage}{0.33\textwidth}
   \begin{center}
    \includegraphics[scale=0.4]{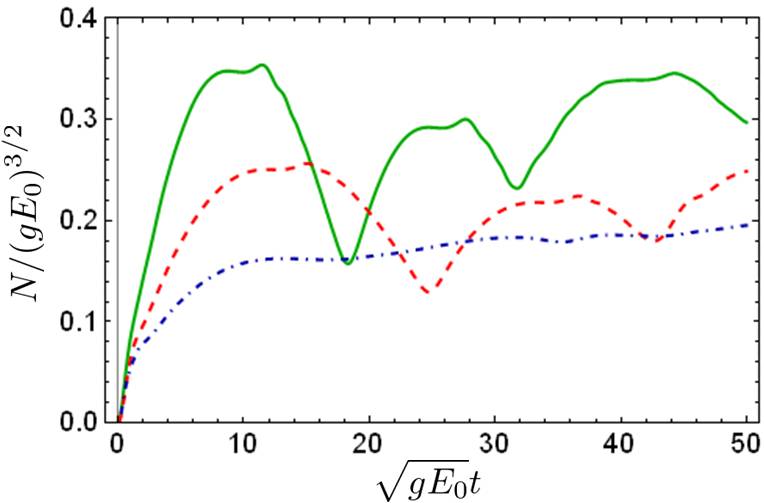} 
   \end{center}
  \end{minipage} \\
  (a) color current density &
  (b) electric field &
  (c) quark number density 
 \end{tabular}
 \caption{Magnetic field dependence. $a=0.01, g=1, \theta =0$.}
 \label{fig:current_mag1}
\end{figure}

\begin{figure}
 \begin{tabular}{ccc}
  \begin{minipage}{0.33\textwidth}
   \begin{center}
    \includegraphics[scale=0.42]{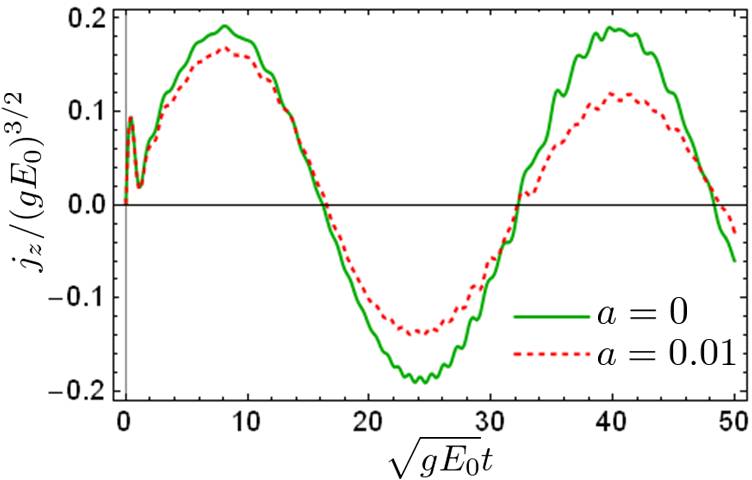} 
   \end{center}
  \end{minipage} & 
  \begin{minipage}{0.33\textwidth}
   \begin{center}
    \includegraphics[scale=0.4]{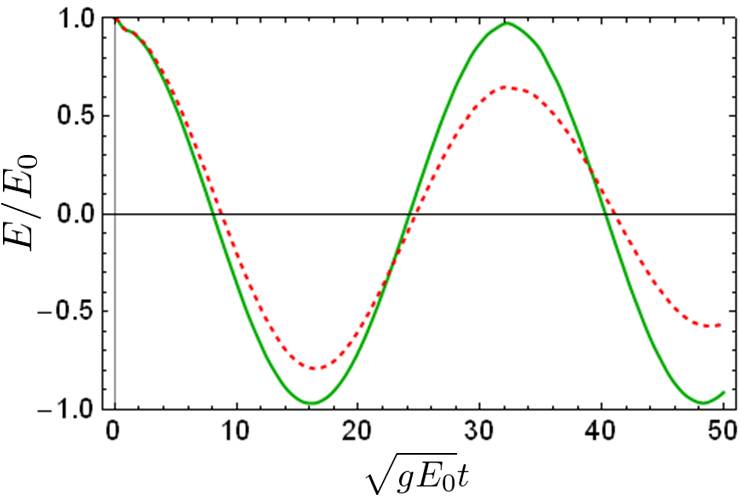} 
   \end{center}
  \end{minipage} &
  \begin{minipage}{0.33\textwidth}
   \begin{center}
    \includegraphics[scale=0.4]{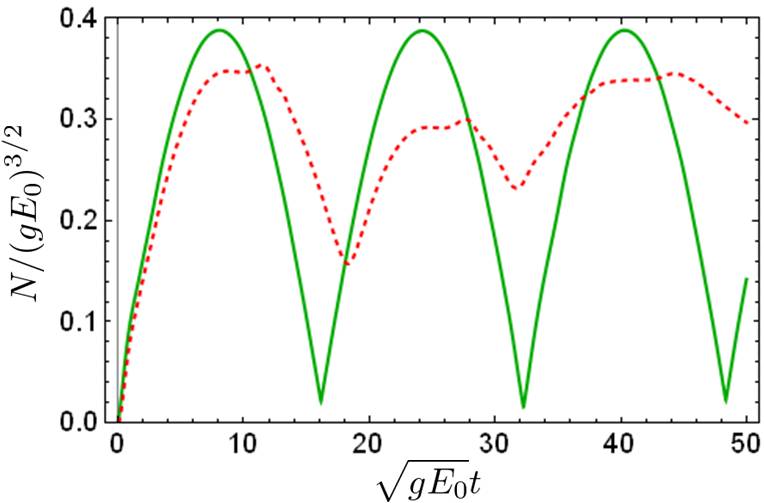} 
   \end{center}
  \end{minipage} \\
  (a) color current density &
  (b) electric field &
  (c) quark number density 
 \end{tabular}
 \caption{Mass dependence under the magnetic field. $B=E_0 , g=1, \theta =0$.}
 \label{fig:current_mag2}
\end{figure}

Fig.\ref{fig:current_mag1} presents the magnetic field dependence of the current, the electric field 
and the particle number density ($a=0.01,g=1,\theta =0$). 
The color current density and the quark number density are indeed enhanced by the magnetic field, 
and as a consequence the frequency of the plasma oscillation increases. 
We can estimate $t_c$, which is a typical time scale of back reaction, in the same way as that in Section \ref{subsec:pp},
and obtain
\begin{align}
t_c = \frac{\pi }{2} \sqrt{\frac{(2\pi )^3 }{4N_f E_0 \sum _i |w_i g|^3 e^{-\frac{\pi m^2}{|w_i |gE_0 }} }
      \frac{\tanh \pi \frac{B}{E_0 }}{\pi \frac{B}{E_0 }} } . \label{t_c inB}
\end{align}
This is a decreasing function of $B$. 

Because only the lowest $m_\text{T}$ mode mainly contributes under a strong magnetic field\footnote{
However, contribution from higher modes is not completely zero if magnetic field strength is finite.
In particular, the peak seen in the current density just after switching on the field does not disappear 
even in the case of massless particle under the magnetic field,
although this is caused by nonzero $m_\text{T}$ modes. }, 
the results in the massless case ($a=0$) and
those with nonzero mass ($a=0.01$) show drastic difference [Fig.\ref{fig:current_mag2}],
whereas in the absence of a magnetic field, there is no considerable difference between them [Fig.\ref{fig:current1}]. 
The particle number density is especially remarkable:
Its time evolution becomes nearly periodic in the massless case.  
Similar behavior of a particle number has been obtained by Iwazaki \cite{Iwazaki2009} in a different way.  
Emergence of this periodic behavior is consistent with our contention 
that the polarization current and the interference of distributions are the source of apparent irreversibility of time,
because as noted in Sections \ref{subsec:damping} and \ref{subsec:interference}, 
the zero mode contributes to neither the polarization current nor the interference term.  

\subsection{Chiral anomaly} \label{subsec:anomaly}
The magnetic field has another effect: induction of chiral charge.   
Because the lowest Landau level ($n=0$) is occupied only by particles with spin parallel to the magnetic field ($s=\uparrow $), 
the balance between spin-$\uparrow$ and $\downarrow$ is broken and thus chiral charge is induced. 

We consider the chiral charge density given by the following equation
\begin{equation}
\mathcal{Q}_5 = \langle 0,\text{in} |:\bar{\psi } \gamma _0 \gamma _5 \psi :|0,\text{in} \rangle . \label{Q5}
\end{equation}
In the chiral limit ($m=0$), the chiral charge equals the difference between the number of right-handed particles
and that of left-handed particles. 

Substituting Eq.\eqref{psi} into Eq.\eqref{Q5} and subtracting a divergent part by the normal ordering,
we can express the chiral charge density in terms of the distribution functions:
\begin{equation} 
\mathcal{Q} _5 (t) = 2N_f \sum _{i=1,2,3} \frac{w_i gB}{2\pi } \int \! \frac{dp_z }{2\pi } \left [
 \frac{p_z }{\omega _{p_z ,n=0 ,\uparrow} } f^i _{p_z , n=0, \uparrow} (t) 
 +\frac{m}{\omega _{p_z ,n=0 ,\uparrow} } g^i _{p_z , n=0, \uparrow} (t) \right ] .
\end{equation}
Note that only the lowest mode contributes to $\mathcal{Q} _5 $. 
Contribution from the higher modes is totally canceled out because they are occupied with both 
the spin-$\uparrow$ and $\downarrow$ modes.
When $m\neq 0$, the anomalous distribution also contributes to the chiral charge.

The time evolution of the chiral charge obeys the Adler-Bell-Jackiw anomaly equation \cite{Adler,Bell-Jackiw}. 
The expectation of the anomaly equation extended to the QCD case now reads\footnote{
The divergence of the chiral current is dropped because of the space homogeneity.}
\begin{equation}
\frac{d}{dt} \mathcal{Q} _5 (t) = \frac{N_f g^2 }{4\pi ^2 } E(t) B +2m \bar{\mathcal{Q}} _5 (t), \label{ABJ}
\end{equation}
where the term $2m \bar{\mathcal{Q}} _5 $ describes explicit breaking of the chiral symmetry due to the mass term, and
\begin{equation}
\begin{split}
\bar{\mathcal{Q}} _5 (t) &= \langle 0,\text{in} |:\bar{\psi } i\gamma _5 \psi :|0,\text{in} \rangle \\
 &= 2N_f \sum _{i=1,2,3} \frac{w_i gB}{2\pi } \int \! \frac{dp_z }{2\pi } 
    \mathrm{Im} \left[ e^{-2i\omega _p t} \alpha ^i _{p_z ,n=0, \uparrow } (t)\beta ^i _{p_z ,n=0, \uparrow } (t) \right] .
\end{split}
\end{equation}

\begin{figure}
 \begin{tabular}{ccc}
  \begin{minipage}{0.36\textwidth}
   \begin{center}
    \includegraphics[scale=0.45]{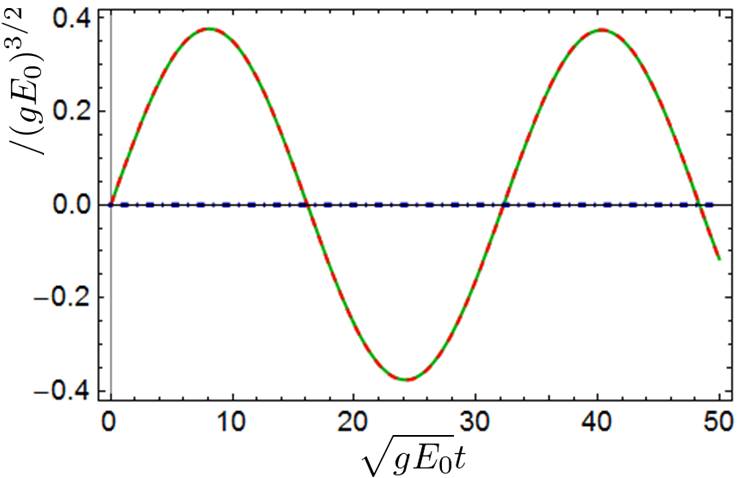} 
   \end{center}
  \end{minipage} & \hspace{-0.5cm}
  \begin{minipage}{0.36\textwidth}
   \begin{center}
    \includegraphics[scale=0.45]{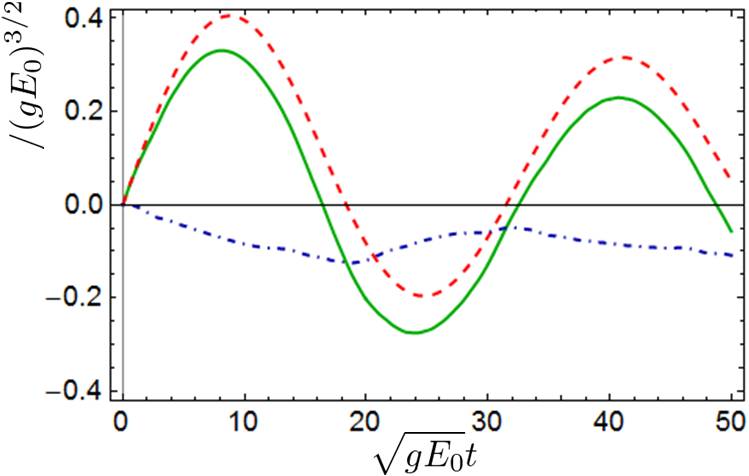} 
   \end{center}
  \end{minipage} & \hspace{-0.5cm}
  \begin{minipage}{0.2\textwidth}
   \begin{center}
    \includegraphics[scale=0.5]{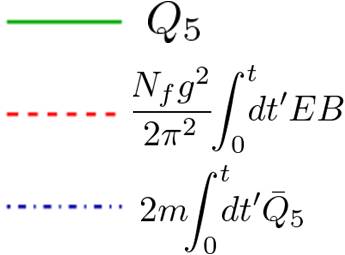} 
   \end{center}
  \end{minipage} \\
  (a)$a=0$ & (b)$a=0.01$ & 
 \end{tabular}
 \caption{Time evolution of the chiral charge. $g=1,B=E_0 ,\theta =0$.}
 \label{fig:chiral}
\end{figure}

In Fig.\ref{fig:chiral}, each term of the time integral of Eq.\eqref{ABJ} is plotted. 
Equation \eqref{ABJ} is indeed satisfied. 
Due to plasma oscillation, the chiral charge density shows oscillation. 
As noted in the previous subsection, time evolution is nearly periodic in the chiral limit [Fig.\ref{fig:chiral}(a)]. 
This curve of $\mathcal{Q} _5$ becomes exactly a sine one in the limit of a strong magnetic field
$B\gg E_0$ \cite{Iwazaki2009}. 
In contrast, if $m\neq 0$, the term $\bar{\mathcal{Q}} _5$ grows, 
and the periodicity seen in $\mathcal{Q} _5$ is destroyed. 

\section{Summary and discussion} \label{sec:summary}
We have studied the time evolution of the system where quantum quark fields 
and a classical color electric field are dynamically coupled with each other. 
By the procedure of Abelianization, we have reduced the equations of motion to the same form as in the Abelian case. 
Effective coupling strength of each quark with the Abelianized electromagnetic field 
depends on the direction of the field in color space. 
Therefore, the longitudinal momentum distributions of created quarks greatly depend on the color direction 
of the gauge field
[Fig.\ref{fig:Ldis2}]. 
However, we have found that the color current density and total particle density are rather insensitive to
the color orientation of the field since those quantities are obtained after summing up all the color components 
[Fig.\ref{fig:current2}]. 

We have obtained the oscillating behavior of the longitudinal momentum distributions, 
and estimated its typical time scale $t_c$ [Eq.\eqref{t_c}] using the empirical analytic expression for 
the quark distributions. 
This $t_c$ also gives a time scale of net particle production. 
The increase of the quark number density is concentrated at initial time and
gets saturated after $t_c$ [Fig.\ref{fig:current1}(c)].
Let us roughly estimate $t_c$ with the parameters motivated by the color glass condensate; $m\simeq 0, N_f =3$ and 
$gE_0 \simeq Q_s ^2 \sim 1 \text{GeV}^2$, where $Q_s$ is a saturation scale.
Then Eq.\eqref{t_c} yields
\begin{align}
t_c \sim 3/g \hspace{5pt} \text{fm}. \label{3/g}
\end{align}
Note in passing that the $\theta $ dependence of $t_c$ is also small, 2\% effect, at most, in the present case. 
This $t_c$ is not drastically small compared with the typical time scale of the initial stage of heavy-ion collisions,
such as thermalization time expected from hydrodynamic calculations $\lesssim 1$ fm (e.g. \cite{Heinz}). 
However, this estimation incorporates only the effect of quark pair creation.  
If gluon pair creation is taken into account, $t_c$ would become much smaller. 
Pair creation of gluons is stronger than that of quarks because (i) gluons are boson, 
so that they are Bose-enhanced rather than Pauli-blocked, 
and (ii) effective coupling of gluons to an electric field is larger than 
that of quarks.\footnote{If one writes a diagram similar to Fig.\ref{fig:triangle} for gluons, a rotating hexagon is obtained,
of which vertices are on a circle with a radius of 1 being bigger than $1/\sqrt{3}$ in the quark case. } 
However, applying the method in the present paper to gluon pair creation, 
we encounter a difficulty of infrared divergence. 
Furthermore, under a magnetic field, one particle energy of a gluon in the lowest Landau level becomes imaginary 
and instability occurs, which is known as the Nielsen--Olesen instability \cite{Nielsen-Olesen,Chang-Weiss}. 
Thus, we need to resolve these difficulties in order to treat both the quark pair and gluon pair creations
in our framework in a unified manner. 

In addition to the plasma oscillation, we have pointed out that the response of vacuum to the electric field involves 
several phenomena such as the Pauli blocking, damping of the electric field and rapid oscillations 
in the quark distribution. 
In particular, it has been revealed that the cause of the rapid oscillations is the interference between the particles. 
Because of this interference, the distribution becomes sensitive to the phase of the Bogoliubov coefficients. 
Furthermore, we have found that apparent irreversibility of time evolution is induced 
by nonzero transverse mass modes through the polarization current and the interference.  

Also the time evolution of pressure of the system has been calculated. 
We have shown that the initial anisotropy in pressure is moderated by pair creation and back reaction
even though our treatment is a mean field one. 
If effects of collisions among particles are taken into account, in other words, if quantum gauge fields are introduced,
the system would be more isotropized. 
It is hoped to investigate these effects, as well as effects of non-uniformity of electromagnetic fields.   

Finally, we have discussed the effects of a magnetic field on quark pair creation. 
We have shown that a magnetic field enhances pair creation because of the emergence of the Landau level 
and the spin-magnetic field interaction. 
As a result, the longitudinal magnetic field speeds up the decay of the electric field. 
This mechanism may have significance in the context of heavy-ion collisions. 

\section*{Acknowledgments}
The author would like to thank Professors T. Matsui and H. Fujii
for enlightening discussions and comments on the manuscript. 
He also acknowledges Professor A. Iwazaki for helpful discussions, 
and Professor T. Hirano for useful comments. 
This work is supported by JSPS research fellowships for Young Scientists.



\begin{thebibliography}{99}

\bibitem{Schwinger}
 J. Schwinger, Phys. Rev. 82 (1951) 664
 
\bibitem{Ruffini}
 R. Ruffini, G. Vereshchagin and S.-S. Xue, Phys. Rept. 487 (2010) 1 [arXiv:0910.0974 [astro-ph.HE]]

\bibitem{Gatoff}
 G. Gatoff, A.K. Kerman and T. Matsui, Phys. Rev. D36 (1987) 114
 
\bibitem{Low}
 F. Low, Phys. Rev. D12 (1975) 163
 
\bibitem{Nussinov}
 S. Nussinov, Phys. Rev. Lett. 34 (1975) 1286

\bibitem{Kovner}
 A. Kovner, L. McLerran and H. Weigert, Phys. Rev. D52 (1995) 6231 [arXiv:hep-ph/9502289];
 Phys. Rev. D52 (1995) 3809 [arXiv:hep-ph/9505320]
 
\bibitem{Lappi}
 T. Lappi and L. McLerran, Nucl. Phys. A772 (2006) 200 [arXiv:hep-ph/0602189]
 
\bibitem{Kharzeev2005}
 D. Kharzeev and K. Tuchin, Nucl. Phys. A753 (2005) 316 [arXiv:hep-ph/0501234]

\bibitem{Kharzeev2007a}
 D. Kharzeev, E. Levin and K. Tuchin, Phys. Rev. C75 (2007) 044903 [arXiv:hep-ph/0602063]
 
\bibitem{Castorina}
 P. Castorina, D. Kharzeev and H. Satz, Eur. Phys. J. C52 (2007) 187 [arXiv:0704.1426 [hep-ph]]

\bibitem{Fukushima2009}
 K. Fukushima, F. Gelis and T. Lappi, Nucl. Phys. A831 (2009) 184 [arXiv:0907.4793 [hep-ph]]
 
\bibitem{Levai-Skokov}
 P. Levai and V. Skokov, arXiv:0909.2323 [hep-ph]

\bibitem{Tanji}
 N. Tanji, Ann. Phys. 324 (2009) 1691 [arXiv:0810.4429 [hep-ph]]

\bibitem{Asakawa}
 M. Asakawa and T. Matsui, Phys. Rev. D43 (1991) 2871
 
\bibitem{Dawson} 
 J.F. Dawson, B. Mihaila and F. Cooper, Phys. Rev. D80 (2009) 014011 [arXiv:0906.2225 [hep-ph]] 

\bibitem{Kluger1991-1993}
 Y. Kluger, J.M. Eisenberg, B. Svetitsky, F. Cooper and E. Mottola,
 Phys. Rev. Lett. 67 (1991) 2427;
 Phys. Rev. D45 (1992) 4659

\bibitem{Kluger1998}
 Y. Kluger, E. Mottola and J.M. Eisenberg, Phys. Rev. D58 (1998) 125015 [arXiv:hep-ph/9803372]

\bibitem{Bloch} 	
 J.C.R. Bloch, V.A. Mizerny, A.V. Prozorkevich, C.D. Roberts, S.M. Schmidt, S.A. Smolyansky and D.V. Vinnik,
 Phys. Rev. D60 (1999) 116011 [arXiv:nucl-th/9907027]

\bibitem{Vinnik} 	
 D.V. Vinnik, A.V. Prozorkevich, S.A. Smolyansky, V.D. Toneev, M.B. Hecht, C.D. Roberts and S.M. Schmidt,
 Eur. Phys. J. C22 (2001) 341 [arXiv:nucl-th/0103073]

\bibitem{Nayak2005a}
 G.C. Nayak and P. Nieuwenhuizen, Phys. Rev. D71 (2005) 125001 [arXiv:hep-ph/0504070]

\bibitem{Nayak2005b}
 G.C. Nayak, Phys. Rev. D72 (2005) 125010 [arXiv:hep-ph/0510052]
 
\bibitem{Ambjorn1979} 
 J. Ambj$\slashchar{\text{o}}$rn, N.K. Nielsen and P. Olesen, Nucl. Phys. B152 (1979) 75

\bibitem{Kharzeev2007b}
 D. Kharzeev, L. McLerran and H. Warringa, Nucl.Phys. A803 (2008) 227 [arXiv:0711.0950 [hep-ph]] 


\bibitem{Ambjorn1983}
 J. Ambj$\slashchar{\text{o}}$rn, J. Greensite and C. Peterson, Nucl. Phys. B221 (1983) 381
 
\bibitem{Iwazaki2009} 
 A. Iwazaki, Phys. Rev. C80 (2009) 052202 [arXiv:0908.4466 [hep-ph]] 
 
\bibitem{Gyulassy-Iwazaki}
 M. Gyulassy and A. Iwazaki, Phys. Lett. 165B (1985) 157

\bibitem{Casher}
 A. Casher, H. Neuberger and S. Nussinov, Phys. Rev. D20 (1979) 179

\bibitem{Habib} 
 S. Habib, Y. Kluger, E. Mottola and J.P. Paz, Phys. Rev. Lett. 76 (1996) 4660 [arXiv:hep-ph/9509413]

\bibitem{Cooper2008} 
 F. Cooper, J.F. Dawson and B. Mihaila, Phys. Rev.D78 (2008) 117901 [arXiv:0811.3905 [hep-ph]]

\bibitem{Adler}
 S.L. Adler, Phys. Rev. 177 (1969) 2426

\bibitem{Bell-Jackiw}
 J.S. Bell and R. Jackiw, Nuovo Cimento A60 (1969) 47

\bibitem{Heinz}
 U.W. Heinz, AIP Conf. Proc. 739 (2005) 163 [arXiv:nucl-th/0407067]
 
\bibitem{Nielsen-Olesen}
 N.K. Nielsen and P. Olesen, Nucl. Phys. B144 (1978) 376

\bibitem{Chang-Weiss}
 S.J. Chang and N. Weiss, Phys. Rev. D20 (1979) 869

\end{thebibliography}
\end{document}